\documentclass[12pt]{article}
\usepackage{ArxivStyle}
\newcommand{\E}{\mathbb{E}}

\usepackage{xcolor}
\definecolor{forestgreen}{RGB}{34,139,34}

\usepackage{subcaption}
\usepackage{pifont}

\title{The Fr\'echet correlation coefficient for heterogeneous random objects}
\author{
Shuaida He\\
School of Computing and Data Science, University of Hong Kong
\and
Yangzhou Chen\thanks{Co-first author.}\\
Department of Statistics and Data Science, Southern University of Science and Technology
\and
Xin Chen\thanks{Corresponding author. Email address: \texttt{chenx8@sustech.edu.cn}.}\\
Department of Statistics and Data Science, Southern University of Science and Technology}
\date{\today}

\begin{document}
\maketitle

\begin{abstract}
    Modern regression analysis often involves responses and predictors taking values in the same or distinct metric spaces. To rank non-Euclidean heterogeneous predictors in regression by explanatory strength, analogous to the classical $R^2$, we introduce the Fr\'echet correlation coefficient (FCC), defined as the relative reduction in the Fr\'echet variance of the response after conditioning on a specific predictor. FCC is directional, model-free,  and interpretable on a unit-scale, attaining one under almost sure functional dependence and zero when the Fr\'echet mean is invariant to conditioning. We propose a novel partition-based estimator that avoids explicit nonparametric estimation of the conditional Fr\'echet mean function, thereby improving both computational efficiency and flexibility. A tailored wild bootstrap algorithm is further developed for testing the Fr\'echet conditional mean dependence. We establish asymptotic theory and evaluate power through extensive simulations. 
\end{abstract}

\section{Introduction}\label{sec:introduction}
Correlation has been a central concept in statistics since the emergence of regression analysis in the late nineteenth century \citep{galton1889co,pearson1896vii}.
In many applications, its value lies not only in detecting association, but also in quantifying explanatory strength on a common, interpretable scale.
For scalar Euclidean variables, this role is exemplified by Pearson's correlation and the coefficient of determination $R^2$: in simple linear regression with an intercept, $R^2$ equals the squared Pearson correlation, so explained variation provides a natural basis for comparing predictors \citep{tjostheim2022statistical}.
Modern data analysis, however, increasingly involves heterogeneous random objects that are more naturally modeled as elements of metric spaces, and predictors and responses may live in different geometries.
In such problems, the central practical question is often directional: how much of the variability of a response $Y$ is explained by a predictor $X$?

This question is especially relevant in multimodal studies with heterogeneous data types. In the Human Connectome Project, for example, a cognitive response constructed from multiple Likert items may be represented as a distribution in a Wasserstein space, while candidate predictors may be Euclidean personality scores, circular sleep variables, compositional behavioral profiles, or brain connectomes represented as symmetric positive definite (SPD) matrices. In such settings, the scientific objective is often to compare heterogeneous predictors by how much variation they explain in a response of interest.
This objective calls for a directional measure of explained variation that respects the geometry of both predictor and response.

A common first strategy is to replace each random object by a convenient Euclidean summary, so that familiar Euclidean coefficients can be applied.
Beyond Pearson's correlation, rank-based coefficients such as Spearman's $\rho$ and Kendall's $\tau$ target monotone association, and more recent proposals such as Chatterjee's correlation \citep{chatterjee2021new} provide directional, model-free summaries of predictive dependence.
In parallel, general-purpose dependence measures such as distance covariance \citep{szekely2007measuring} and kernel-based criteria \citep{gretton2005measuring} yield powerful omnibus procedures for testing departures from independence in Euclidean settings.
The difficulty is that scalar or Euclidean reductions may discard precisely the aspect of the response variation that is scientifically relevant. As a result, predictor rankings based on coarse summaries can be unstable or misleading when the response is intrinsically non-Euclidean or when different geometries encode different scientific questions.

Recent dependence measures for random objects in metric spaces, including ball covariance \citep{pan2020ball,wang2024nonparametric} and profile association \citep{zhou2025association}, provide valuable tools for detecting association, but they target a different inferential goal.
These measures are typically symmetric in $(X,Y)$ and are not defined through a variance decomposition of the response, so they do not directly yield a directional $R^2$-type effect size for response-specific explained variation.
This is a genuine difference: in general metric spaces, there is typically no linear projection identity, conditional Fr\'echet means can be difficult to estimate, and even defining explained variation requires care when predictor and response lie in different geometries.

These considerations motivate the central problem of this paper: given a predictor $X$ taking values in a metric space $(\mathcal X,d_X)$ and a response $Y$ taking values in a possibly different metric space $(\mathcal Y,d_Y)$, can we define a directional, scale-free, and interpretable coefficient that quantifies the proportion of Fr\'echet variation of $Y$ explained by $X$,  enables comparison of heterogeneous predictors on a common scale, and can be estimated in a convenient, model-free way?

We answer this question by introducing the Fr\'echet correlation coefficient (FCC), an explained variation functional defined by the reduction in Fr\'echet variance after conditioning on the predictor. Under mild regularity conditions, FCC is well defined even when $X$ and $Y$ take values in different metric spaces.
At the population level, FCC equals one if and only if $Y$ is almost surely a measurable function of $X$, and it equals zero if and only if the conditional and unconditional Fr\'echet means coincide almost surely. Independence therefore implies FCC equals zero, but the converse need not hold.
This distinction is intentional: FCC is designed as a directional explained-variation coefficient, so its null hypothesis targets the absence of Fr\'echet mean dependence rather than all possible forms of dependence.  At the sample level, we propose an efficient partition-based estimator that avoids explicit nonparametric estimation of the conditional Fr\'echet mean function. The resulting procedure is straightforward to implement, and well suited to comparing heterogeneous predictors in metric-space settings.

We introduce the following example to contrast the classical Euclidean notion of explained variation with the metric-space notion captured by FCC.  It highlights the key gap addressed here: for structured random-object responses, explained variation is inherently geometry dependent.
\begin{example}[Washington bike sharing]\label{eg1}
Bike-sharing data provide a transparent setting in which different response geometries correspond to different scientific questions. Using the 2011--2012 Washington Capital Bikeshare data for illustration \citep{fanaee2014event}, suppress the day index and let $C(h)$ denote the number of rentals in hour $h\in\{0,\ldots,23\}$ on a given day.
If the goal is to quantify overall demand, a natural Euclidean response is the total number of daily rentals,
\[
    Y=\sum_{h=0}^{23} C(h)\in\mathbb{R}.
\]
For a scalar Euclidean predictor $X$, a standard summary is the linear explained-variation coefficient $R^2(Y,X)$, which in simple linear regression coincides with the squared Pearson correlation.
Instead, if the scientific question concerns \textit{when} rentals occur within a day, a more appropriate response is the normalized demand profile
\[
    \widetilde{Y}(h)=\frac{C(h)}{\sum_{k=0}^{23}C(k)},\qquad h=0,\ldots,23,
\]
which can be viewed as a probability distribution on the ordered hour support, that is, an element of $\mathcal{W}_2(\{0,\ldots,23\})$. This representation retains timing and shape information that is lost when the day is reduced to a single scalar total.
Accordingly, the relevant directional explained-variation summary is FCC, namely $\rho(\widetilde{Y},X)$, rather than a classical Euclidean $R^2$.

Figure~\ref{fig:bike} compares several representative predictors under these two response targets. Mean temperature and working-day status are evaluated both for total rentals and for the normalized daily profile. Humidity is represented either by a scalar daily average or by the full within-day trajectory, and weekly timing is encoded either by a linear day-of-week code or by circular phase. Finally, within-day weather dependence is summarized by an SPD object, which has no natural scalar Euclidean analogue.
The purpose here is not to argue that FCC should numerically dominate Euclidean $R^2$, but to emphasize that the two coefficients quantify different aspects of explanatory strength.
In this example, working-day status has negligible Euclidean $R^2$ for total rentals but a substantially larger FCC for the daily profile, suggesting that it primarily explains \textit{when} rentals occur rather than \textit{how many} occur in total.
\begin{figure}[!ht]
	\centering
	\begin{subfigure}[t]{0.4\textwidth}
		\centering
		\includegraphics[width=\linewidth]{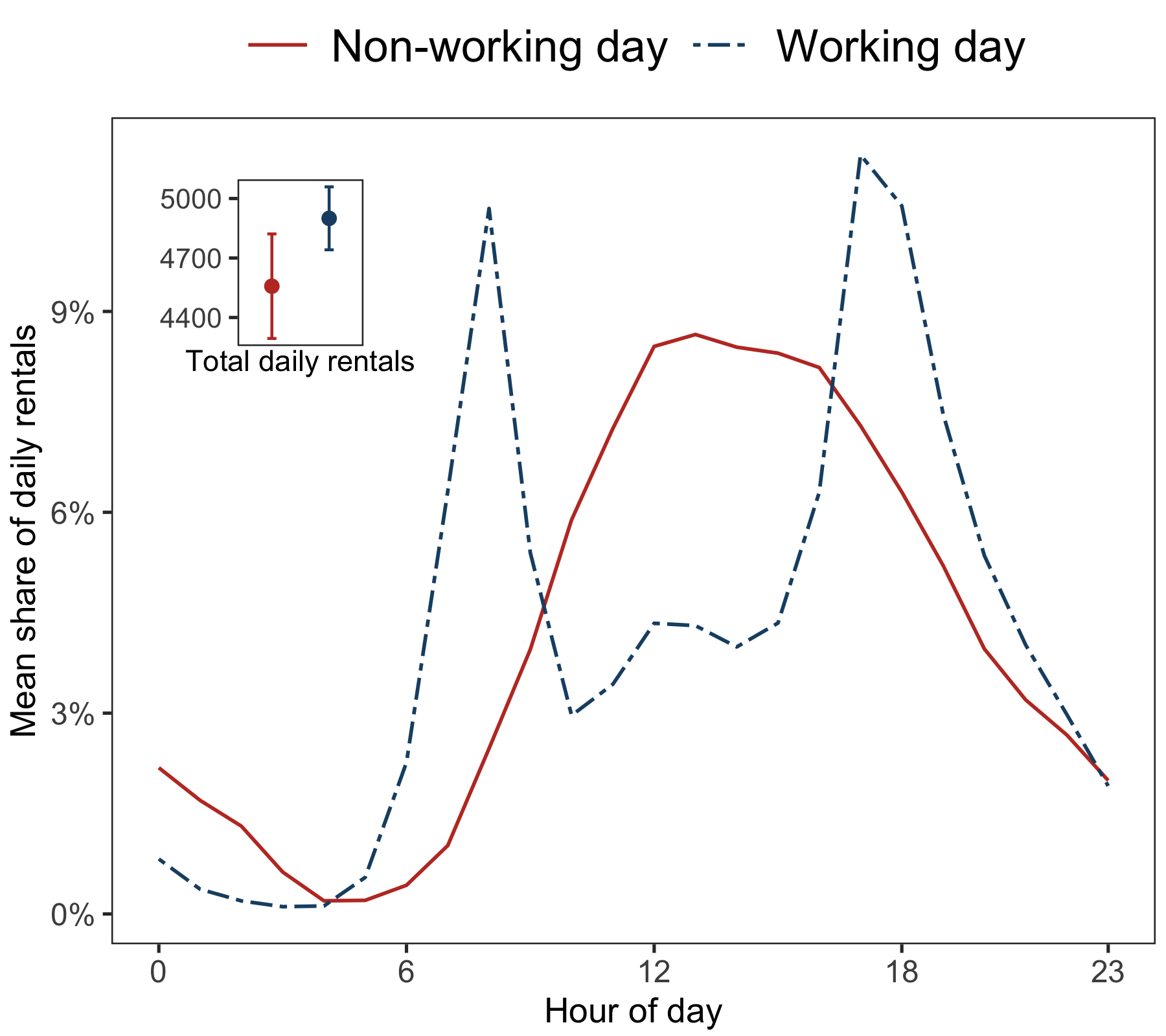}
		\caption{Normalized rental demand profiles versus total-demand insets.}
	\end{subfigure}\hfill
	\begin{subfigure}[t]{0.58\textwidth}
		\centering
		\includegraphics[width=\linewidth]{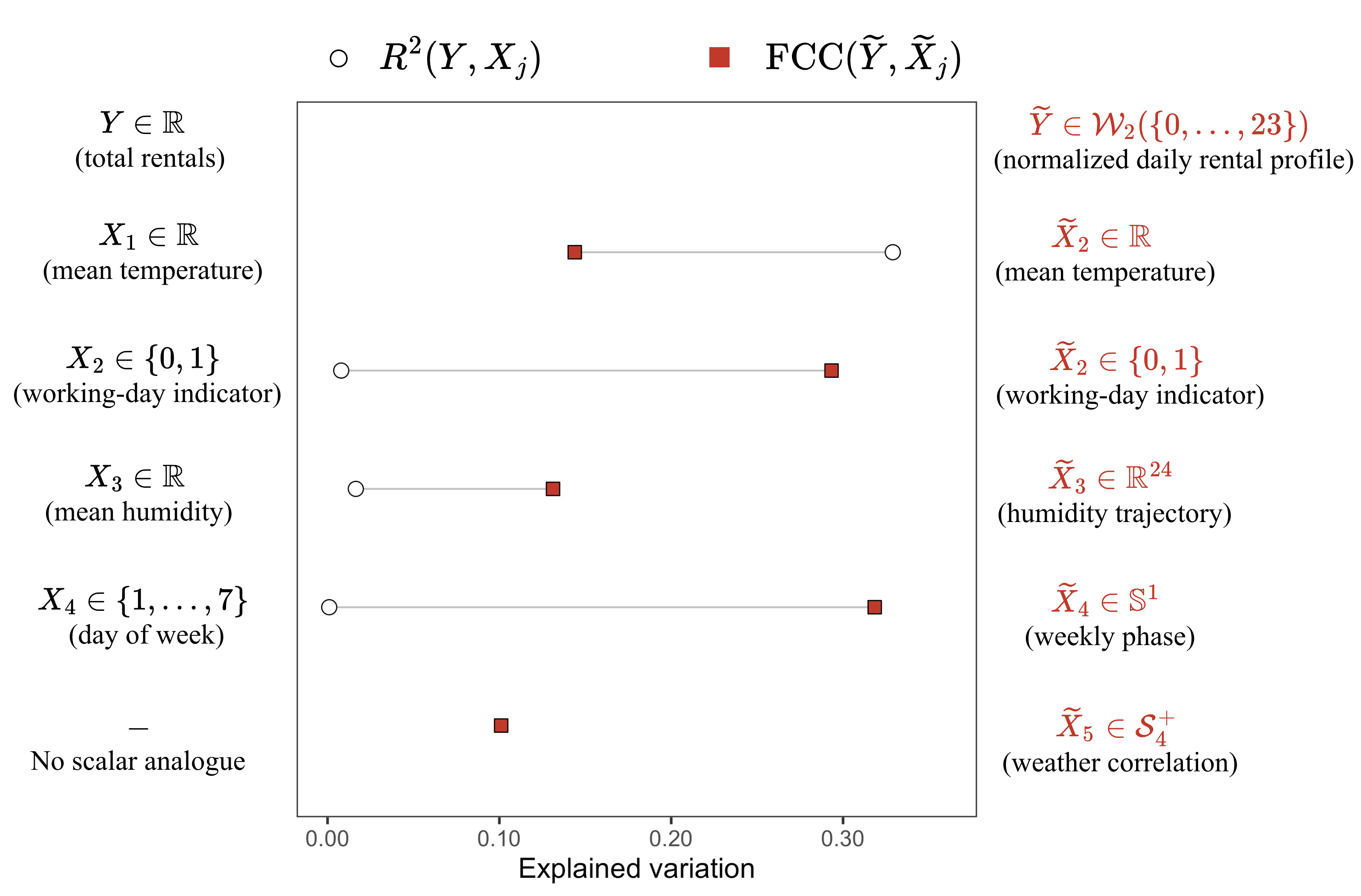}
		\caption{Euclidean $R^2$ versus metric FCC.}
	\end{subfigure}
	\caption{Washington bike-sharing example. Panel (a) plots mean normalized daily rental profiles $\widetilde{Y}$ for working and non-working days, with an inset showing mean total daily rentals $Y$ and $95\%$ confidence intervals. Working days exhibit clear morning and evening commute peaks, whereas non-working days peak around midday; the difference in total daily rentals is comparatively modest. Panel (b) compares Euclidean summaries for total daily rentals $Y$, reported through $R^2$, with metric-space summaries for the normalized daily profile $\widetilde{Y}$, reported through FCC. The selected rows illustrate the same predictor under different response targets, alternative geometric representations of the same covariate, and a metric predictor without a natural scalar Euclidean analogue.}
	\label{fig:bike}
\end{figure}
\end{example}

FCC connects to several established lines of work. When both the response and covariate are scalar Euclidean variables, it reduces to the well-known generalized measure of correlation of \citet{zheng2012generalized}.
It is also in the spirit of Fr\'echet analysis of variance for random objects \citep{dubey2019frechet}, but shifts the focus from group comparison to response-specific explained variation.
FCC further complements the seminal work Fr\'echet regression \citep{petersen2019frechet} and its extensions \citep{bhattacharjee2023single,ying2022frechet,zhang2024dimension} by providing a generic summary of regression strength even when the predictor and response need not share the same geometry.
In particular, FCC is related to the Fr\'echet coefficient of determination \(R_\oplus^2\) of \citet{petersen2019frechet}, but targets the full conditional Fr\'echet mean rather than a specified global Fr\'echet regression model with Euclidean predictors; consequently, FCC is model-free, applies to predictors taking values in metric spaces, and satisfies \(R_\oplus^2\le \rho\) whenever both quantities are defined, with equality under correct specification of the global Fr\'echet regression function.
Finally, FCC is complementary to metric distribution function derived association measures \citep{wang2024nonparametric}, which pursue full distributional dependence via product-space characterizations, whereas FCC focuses on directional explained variation through conditional Fr\'echet means and therefore does not require full-law reconstruction.

The contributions of this paper are threefold.
First, we formulate FCC as a directional explained-variation functional for random objects in metric spaces and establish its population-level interpretation, including sharp characterizations of the boundary cases FCC$=0$ and FCC$=1$.
Second, we propose an efficient partition-based estimator,  prove its consistency, and derive null asymptotic theory in two regimes: a fixed-partition regime with a weighted chi-square limit and a diverging-partition regime with either Gaussian or weighted chi-square limits, depending on the cumulative variance structure.
Third, we study representative Wasserstein and SPD settings, and use simulations to show that FCC yields an interpretable way to compare heterogeneous predictors by explanatory strength.
Together, these results position FCC as a practical correlation coefficient for heterogeneous non-Euclidean data.

\section{Methodology}\label{sec:method}
\subsection{Preliminaries}\label{subsec:Preliminaries}
Let $(\Omega,\mathcal{F},\mathbb{P})$ be a probability space. Let $X:\Omega\to(\mathcal X,d_X)$ and $Y:\Omega\to(\mathcal Y,d_Y)$ be random elements, where $(\mathcal X,d_X)$ and $(\mathcal Y,d_Y)$ are bounded, complete, separable metric spaces.
Write $P_X:=\mathbb P\circ X^{-1}$ for the marginal law of $X$ on $\mathcal X$.

For a random element taking values in a metric space, the Fr\'echet mean and variance generalize the classical mean and variance \citep{frechet1948elements,dubey2019frechet}.
Specifically, the Fr\'echet mean of $Y$ is any minimizer of $\mu\mapsto\mathbb E\,d_Y^2(Y,\mu)$ over $\mu\in\mathcal Y$. When this minimizer is unique, we denote it by
\[
	\mu_F:=\arg\min_{\mu\in\mathcal Y}\mathbb E\,d_Y^2(Y,\mu),
\]
and define the corresponding Fr\'echet variance by
\[
	V_F:=\mathbb E\,d_Y^2(Y,\mu_F).
\]

Since $\mathcal X$ and $\mathcal Y$ are complete separable metric spaces, a regular conditional distribution of $Y$ given $X$ exists. Fix a version and denote it by $P_{Y\mid X}(\cdot\mid x)$.
For notational convenience, throughout we interpret $\mathbb E[\cdot\mid X=x]$ with respect to $P_{Y\mid X}(\cdot\mid x)$ for $P_X$-almost every $x\in\mathcal X$, and define the conditional Fr\'echet mean by
\[
	\mu_{F,x}:=\arg\min_{\mu\in\mathcal Y}\int_{\mathcal Y} d_Y^2(y,\mu)\,P_{Y\mid X}(dy\mid x),
\]
and the associated conditional Fr\'echet variance by
\[
	V(x):=\mathbb E \bigl[d_Y^2(Y,\mu_{F,x})\mid X=x\bigr].
\]
We write $\mu_{F,X}$ and $V(X)$ for the corresponding $\sigma(X)$-measurable versions.

\subsection{The proposed metric}\label{subsec:The proposed metric}
We define the Fr\'echet correlation coefficient (FCC) by
\[
	\rho=1-\frac{\mathbb E[V(X)]}{V_F}.
\]
The quantity $\mathbb E[V(X)]$ is the residual Fr\'echet variation of $Y$ after conditioning on $X$, so $\rho$ measures the proportion of Fr\'echet variation in $Y$ explained by $X$.
In general, the explained variation of $Y$ by $X$ need not coincide with the explained variation of $X$ by $Y$.
Thus, FCC is a directional explained-variation coefficient for the response $Y$ given the predictor $X$.

The following assumptions ensure that FCC is well-defined.
\begin{assumption}\label{assumption:mean-existence}
	The global Fr\'echet mean $\mu_F$ exists and is unique, and the conditional Fr\'echet mean $\mu_{F,x}$ exists and is unique for $P_X$-almost every $x \in \mathcal{X}$.
\end{assumption}

\begin{assumption}\label{assumption:nonzero_variance}
	The Fr\'echet variance of $Y$ is strictly positive, that is, $V_F > 0.$
\end{assumption}
Assumpton~\ref{assumption:mean-existence} ensures that both the global Fr\'echet variance $V_F$ and the residual term $\mathbb E\{V(X)\}$ are well defined. Assumpton~\ref{assumption:nonzero_variance} is the exact nondegeneracy condition needed to define $\rho$, which holds whenever $Y$ is not almost surely constant under the metric $d_Y$. Together, they ensure $\rho$ is properly behaved.
We are ready to vist its sample estimation.

\subsection{Partition-based estimator}\label{subsec:Sample estimation}
Direct estimation of $\mathbb E\{V(X)\}$ would require recovering the full conditional Fr\'echet mean map $x\mapsto\mu_{F,x}$, which is a difficult nonparametric problem on a general metric space. We therefore suggest a partition-based approximation. Let $\{\Omega_m\}_{m=1}^{M_n}$ be a measurable partition of $\mathcal X$; when this partition depends on $n$, we suppress that dependence in the notation. For each cell, define the population cellwise Fr\'echet mean and variance by
\[
	\mu_{F,m}:=\arg\min_{\mu\in\mathcal Y}\mathbb E\left[d_Y^2(Y,\mu)\mid X\in\Omega_m\right],
\qquad
V_m:=\mathbb E\left[d_Y^2(Y,\mu_{F,m})\mid X\in\Omega_m\right].
\]
Then $\sum_{m=1}^{M_n}\Pr(X\in\Omega_m)V_m$ is a partition-based approximation to $\mathbb E\{V(X)\}$. The shrinking-mesh and growth conditions needed for consistency and null asymptotics will be imposed later, when those results are stated.

Given i.i.d.\ observations $\{(X_i,Y_i)\}_{i=1}^n$, let $n_m=\sum_{i=1}^n \mathbf{1}\{X_i\in\Omega_m\}$. Define the global and cellwise sample Fr\'echet means by
\[
\hat\mu_F := \arg\min_{\mu\in\mathcal Y}\frac1n\sum_{i=1}^{n} d_Y^2(Y_i,\mu),
\qquad
\hat\mu_{F,m} := \arg\min_{\mu\in\mathcal Y}\frac1{n_m}\sum_{i:\,X_i\in\Omega_m} d_Y^2(Y_i,\mu),
\]
and the corresponding sample Fr\'echet variances by
\[
\hat V_F := \frac1n\sum_{i=1}^{n} d_Y^2\left(Y_i,\hat\mu_F\right),
\qquad
\hat V_m := \frac1{n_m}\sum_{i:\,X_i\in\Omega_m} d_Y^2\left(Y_i,\hat\mu_{F,m}\right).
\]
With these quantities in place, we estimate $\rho$ by
\[
		\rho_n^{M_n}:= 1 - \frac{\sum_{m=1}^{M_n}\frac{n_m}{n}\hat V_m}{\hat V_F}.
\]
This estimator is defined on the event $\{\min_{1\le m\le M_n}n_m>0\}$. A cell-probability condition introduced in Section~\ref{subsec:Asymptotic consistency} guarantees that this event has probability tending to one in the asymptotic regimes considered in Section~\ref{sec:theory}.

{\subsection{Partition construction}}\label{subsec:partition-construction}
The estimator $\rho_n^{M_n}$ is defined for any measurable partition $\{\Omega_m\}_{m=1}^{M_n}$ of the predictor space $\mathcal X$. Note that the partition should be chosen using the geometry of the predictor space only. This choice keeps FCC model-free on the response side and avoids response-adaptive partitioning, which would otherwise complicate both consistency and inference.

Several constructions are natural in practice. If $X$ is discrete, or if scientifically meaningful groups are available a priori, one can take those groups as the partition cells.
For Euclidean predictors, quantile binning and recursive axis-aligned splits provide simple defaults.
For general metric-space predictors, however, it is preferable to use coordinate-free constructions that depend only on $d_X$.

A particularly simple option is a prototype-based random partition.
Let $\xi_1,\dots,\xi_{M_n}\in\mathcal X$ be prototype points chosen independently of the estimation sample, either from a reference distribution on $\mathcal X$ or from an auxiliary predictor sample.
With a deterministic tie-breaking rule, define the nearest-prototype (Voronoi) cells \citep{aurenhammer1991voronoi}
\[
	\Omega_m:=\Bigl\{x\in\mathcal X:\ d_X(x,\xi_m)\le d_X(x,\xi_k)\ \text{for all }k=1,\dots,M_n\Bigr\},\qquad m=1,\dots,M_n.
\]
This yields a measurable partition of $\mathcal X$ and supports out-of-sample assignment for new predictor values.

Another approach is clustering \citep{chaudhuri2014rates}. Using predictors only, one may obtain representative prototypes $c_1,\dots,c_{M_n}$ via a metric clustering method such as $k$-medoids, and then define Voronoi cells around the fitted prototypes. For a new predictor value $x$, cell membership is assigned by nearest-prototype classification under $d_X$. In this way the partition remains well defined on the full predictor space.

From a theoretical standpoint, what matters is not the specific partitioning algorithm but the resulting cell geometry. The conditions required below are that each cell carries sufficient probability mass and that the partition becomes increasingly fine ({see the assumptions of Proposition~\ref{pro:consistency}}). Random and clustering-based partitions are therefore both admissible, provided these conditions hold with high probability.

When the partition is learned from data, it is preferable to construct it using an auxiliary sample, or at least using the predictor sample alone, and then evaluate $\rho_n^{M_n}$ on an independent sample.
This data-splitting strategy keeps the partition exogenous to the estimating sample and yields a clean extension of the consistency theory in Section~\ref{subsec:Asymptotic consistency}.
In practice, the efficiency loss from splitting can be reduced by cross-fitting.

\section{Theoretical Analysis}\label{sec:theory}
This section develops the main theoretical properties of FCC. We first establish basic bounds and characterize the boundary cases $\rho=0$ and $\rho=1$, clarifying how the scale should be interpreted in applications. We then prove consistency of the partition-based estimator and derive its null limits in both fixed-partition and growing-partition regimes. Finally, we study the effect of noise on FCC in representative Wasserstein and SPD models.

\subsection{Basic properties}\label{subsec:Characterization}
We begin by showing that FCC always lies in the unit interval. This justifies interpreting it as a standardized explained-variation coefficient for the response $Y$ given the predictor $X$.

\begin{proposition}\label{prop:rho-bounds}
	Under Assumptions~\ref{assumption:mean-existence} and \ref{assumption:nonzero_variance}, the correlation coefficient $\rho$ satisfies $0 \le \rho \le 1.$
\end{proposition}

The two boundary values admit natural interpretations. The case $\rho=1$ corresponds to perfect predictability of $Y$ from $X$ in regression analysis, whereas $\rho=0$ corresponds to the absence of Fr\'echet mean dependence. The next two theorems formalize these statements.

\begin{theorem}\label{thm:rho-equals-1}
	Under Assumptions~\ref{assumption:mean-existence} and \ref{assumption:nonzero_variance}, the coefficient $\rho$ attains its maximum value $\rho = 1$ if and only if there exists a measurable map $f:\mathcal X\to\mathcal Y$ such that
	\[
		Y = f(X) \quad \text{almost surely}.
	\]
	In that case, $f(X)=\mu_{F,X}$ almost surely.
\end{theorem}
Equivalently, $\rho=1$ holds exactly when $V(X)=0$ almost surely, so conditioning on $X$ removes all Fr\'echet variation in $Y$.

\begin{theorem}\label{thm:rho-equals-0}
	Under Assumptions~\ref{assumption:mean-existence} and \ref{assumption:nonzero_variance}, $\rho=0$ if and only if $\mu_{F,X}=\mu_{F}$ almost surely. In particular, if $X$ and $Y$ are independent, then $\rho=0$.
\end{theorem}
Taken together, Proposition~\ref{prop:rho-bounds} and Theorems~\ref{thm:rho-equals-1}--\ref{thm:rho-equals-0} show that FCC behaves as a generalized $R^2$ in metric spaces: it is zero when conditioning on $X$ does not change the Fr\'echet mean of $Y$, and it equals one when $Y$ is completely determined by $X$.

\subsection{Consistency of the partition estimator}\label{subsec:Asymptotic consistency}
We next study the large-sample behavior of $\rho_n^{M_n}$. To state the consistency result, we first impose regularity conditions on the partition and on the cellwise Fr\'echet means.

\begin{assumption}\label{assumption:cell-prob}
    The partition $\{\Omega_m\}_{m=1}^{M_n}$ is balanced in the sense that $p_m := \Pr(X \in \Omega_m) \asymp M_n^{-1}$ uniformly over $1\le m\le M_n$; that is, there exist constants $0<c_1\le c_2<\infty$ independent of $n$ and $m$ such that $c_1/M_n \le p_m \le c_2/M_n$ for all $1\le m\le M_n$.
\end{assumption}
This condition imposes a balanced partition condition, analogous to the sliced stability condition \citep{lin2018consistency} used in Euclidean settings. It ensures asymptotically sufficient cell counts (Lemma~\ref{lemma:n_m-order}), so empty or sparse cells are negligible and cellwise Fr\'echet means and variances can be estimated reliably.

\begin{assumption}\label{assumption:unique-geodesic}
	For $\mathbb{P}$-almost every $y \in \mathcal{Y}$, the map
	$\mu \mapsto d_Y^{2}(\mu,y)$ is geodesically convex on $\mathcal{Y}$. Furthermore, the cellwise Fr\'echet mean $\mu_{F,m} := \arg\min_{\mu \in \mathcal{Y}}
	\mathbb{E}[d_Y^2(Y,\mu)\mid X \in \Omega_m]$ exists and is unique for every partition cell $\Omega_m$.
\end{assumption}

Assumptions~\ref{assumption:cell-prob} and \ref{assumption:unique-geodesic}, together with the boundedness of $\mathcal Y$ and the Lipschitz continuity of $d_Y^2(\cdot,y)$, ensure that the conditions of Theorem~2 in \citet{brunel2023geodesically} are satisfied. Consequently, the sample Fr\'echet means $\hat\mu_F$ and $\hat\mu_{F,m}$ converge to their population counterparts, which yields consistency of the partition-based estimator.

\begin{remark}\label{remark:consistency}
For consistency, Assumption~\ref{assumption:unique-geodesic} is sufficient but need not to be necessary. By \citet{dubey2019frechet}, uniqueness of the Fr\'echet mean together with the following separation condition is also sufficient: for any $\varepsilon>0$,
	\[
		\inf_{d_Y(z,\mu_{F})>\varepsilon} \mathbb{E}\left[ d^2_Y(z,Y)  \right] >\mathbb{E}\left[ d^2_Y(\mu_F,Y)  \right],
	\]
	\[
		\inf_{d_Y(z,\mu_{F,m})>\varepsilon} \mathbb{E}\left[ d^2_Y(z,Y)  \mid X\in\Omega_m \right] >\mathbb{E}\left[ d^2_Y(\mu_{F,m},Y)  \mid X\in\Omega_m \right],\quad \forall m.
	\]
	Under this alternative condition, the sample Fr\'echet means still converge in probability to their population counterparts, which is enough to obtain consistency of $\rho_n^{M_n}$.
\end{remark}

Under these conditions, the following proposition shows that the weighted cellwise sample Fr\'echet variances provide a consistent estimate of the residual variation term $\mathbb E\{V(X)\}$.
\begin{proposition}\label{pro:consistency}
	Under Assumptions~\ref{assumption:mean-existence},~\ref{assumption:cell-prob} and \ref{assumption:unique-geodesic}, suppose in addition that
	\[
	\sup_{1\le m\le M_n}\sup_{x\in\Omega_m} d_Y(\mu_{F,m},\mu_{F,x})\to 0.
	\]
	Then
	\[
	\sum_{m=1}^{M_n}\frac{n_m}{n}\,\hat V_m-\mathbb{E}\{V(X)\}=o_P(1).
	\]
\end{proposition}

Combining Proposition~\ref{pro:consistency} with the consistency of the global Fr\'echet variance estimator $\hat V_F$, we obtain the following result.

\begin{theorem}\label{thm:consistency}
	Under Assumptions~\ref{assumption:mean-existence},~\ref{assumption:cell-prob}, and \ref{assumption:unique-geodesic}, and if in addition
	\[
	\sup_{1\le m\le M_n}\sup_{x\in\Omega_m} d_Y(\mu_{F,m},\mu_{F,x})\to 0,
	\]
	then
	\[
	\rho_n^{M_n}-\rho=o_P(1).
	\]
\end{theorem}
Theorem~\ref{thm:consistency} establishes $\rho_n^{M_n}$ is a consistent estimator of the population FCC. We next study its null asymptotic distribution for the hypothesis of no Fr\'echet mean dependence,
\begin{equation}\label{eq:null}
    H_0:\mu_{F,X}=\mu_F \quad \text{a.s}.
\end{equation}	
which is equivalent to $H_0:\rho=0$ by Theorem~\ref{thm:rho-equals-0}.

\subsection{Asymptotic distribution under the null hypothesis}\label{subsec:Asymptotic distribution}
The form of the null limit depends on both the geometry of $\mathcal Y$ and the asymptotic regime of the partition. We first treat finite-dimensional Riemannian manifolds, where local expansions of the squared geodesic distance are available, and then one-dimensional Wasserstein responses, where the argument proceeds through the $L^2$ quantile embedding.

\subsubsection{Null asymptotics on a finite-dimensional Riemannian manifold}\label{subsubsec:null-riemann}
We first treat the finite-dimensional Riemannian-manifold case. In addition to Assumptions~\ref{assumption:mean-existence}, \ref{assumption:cell-prob}, and \ref{assumption:unique-geodesic}, we impose the following smoothness and nondegeneracy condition.

\begin{assumption}
	\label{assumption:CLT}
	Suppose that $\mathcal Y$ is a finite-dimensional complete Riemannian manifold and that there exists $\delta>0$ such that, for $\mathbb P$-almost every $Y$, the map
	$y\mapsto d_Y^2(Y,y)$ is three times continuously differentiable on
	$\{y:d_Y(y,\mu_F)<\delta\}$, and for each $m$, for $\mathbb P(\cdot\mid X\in\Omega_m)$-almost every $Y$,
	the map $y\mapsto d_Y^2(Y,y)$ is three times continuously differentiable on
	$\{y:d_Y(y,\mu_{F,m})<\delta\}$.
	Moreover, the Hessian operators
	\[
		\mathbb E\!\left[\nabla^2 d_Y^2(Y,\mu_F)\right]
		\quad\text{and}\quad
		\mathbb E\!\left[\nabla^2 d_Y^2(Y,\mu_{F,m})\mid X\in\Omega_m\right]
	\]
	are strictly positive definite, and we assume that the second- and third-order derivatives satisfy the uniform boundedness condition:
	\[
		\sup_{d_Y(y,\mu_F)<\delta}\left\|\nabla^2 d_Y^2(Y,y)\right\| < \infty,
		\quad
		\sup_{d_Y(y,\mu_F)<\delta}\left\|\nabla^3 d_Y^2(Y,y)\right\| < \infty,
	\]
	and
	\[
		\sup_{1 \le m \le M_n}\sup_{d_Y(y,\mu_{F,m})<\delta}\left\|\nabla^2 d_Y^2(Y,y)\right\| < \infty,
		\quad
		\sup_{1 \le m \le M_n}\sup_{d_Y(y,\mu_{F,m})<\delta}\left\|\nabla^3 d_Y^2(Y,y)\right\| < \infty,
	\]
	for some $\delta > 0$.
\end{assumption}

These smoothness and nondegeneracy conditions control the Taylor remainder and justify the local quadratic expansions used in the manifold-based null asymptotic analysis. We distinguish two regimes according to the complexity of the partition.

\medskip
\noindent\textbf{Case 1: $M_n = O(1)$.}
Here the partition is treated as fixed in advance. This regime is appropriate when the cells represent a prespecified coarse discretization of $\mathcal X$, and the null limit reflects the between-cell fluctuations induced by that discretization.

\medskip
\noindent\textbf{Case 2: $M_n \asymp n^{a}$ with $0<a<1/3$.}
Here the partition becomes increasingly fine with $n$. The growth restriction balances approximation accuracy against within-cell sample sizes and yields a nondegenerate large-sample limit while keeping enough observations in each cell for stable estimation.

We begin with a lemma controlling the cell sample sizes. In particular, it guarantees that for each fixed cell index, the within-cell sample size diverges almost surely.

\begin{lemma}\label{lemma:n_m-order}
	Suppose Assumption~\ref{assumption:cell-prob} holds and \(M_n \asymp n^{a}\) for some \(a\in[0,1/3)\). Then, for any fixed \(\delta\in(0,1)\), 
	\(\Pr\{\max_{1\le m\le M_n}|n_m/(np_m)-1|>\delta\}\to0\). Consequently, 
	\(\min_{1\le m\le M_n} n_m \asymp_P n/M_n\asymp_P n^{1-a}\) and 
	\(\max_{1\le m\le M_n} n_m \asymp_P n/M_n\asymp_P n^{1-a}\). 
	In particular, uniformly over \(1\le m\le M_n\), \(n_m=O_P(n^{1-a})\). 
	Moreover, \(\min_{1\le m\le M_n}n_m\to\infty\) almost surely as \(n\to\infty\).
\end{lemma}

\begin{remark}
	\label{remark:n_m-order}
	Lemma~\ref{lemma:n_m-order} shows that
	\(\min_{1\le m\le M_n} n_m\to\infty\) almost surely.
	Therefore, in subsequent asymptotic arguments involving the growing partition, we may work on the event that all cells are eventually nonempty and that the minimum cell size diverges. This event has probability one and hence does not affect the limiting results.
\end{remark}

The next proposition establishes a central limit theorem for $\sum_{i=1}^{n}\left\{ d_Y^2(Y_i,\hat\mu_F)-d_Y^2(Y_i,\mu_F)\right\rbrace$ and $\sum_{\{i:X_i\in\Omega_m\}}\left\{ d_Y^{2}(Y_i,\hat{\mu}_{F,m})- d_Y^{2}(Y_i,{\mu}_{F,m})\right\rbrace$ ($\forall m$).
It provides the key tool needed for the asymptotic distribution of our statistic.

\begin{proposition}\label{pro:CLT}
	Under Assumptions~\ref{assumption:mean-existence}, and \ref{assumption:cell-prob}–\ref{assumption:CLT}, we have
	\[
	\sum_{i=1}^{n}\left\{ d_Y^2(Y_i,\hat\mu_F)-d_Y^2(Y_i,\mu_F)\right\rbrace
	\Rightarrow
	-\frac{1}{2}Z^\top\Lambda^{-1}Z,
	\]
	where $Z\sim\mathcal N(\mathbf 0,C)$, $\Lambda=HF(\mu_F)$ and
	$C=\mathrm{Var}(\psi(Y))$ with $\psi(Y)=2\log_{\mu_F}(Y)$.

	Moreover, for each $m$, we have
	\[
	\sum_{\{i:X_i\in\Omega_m\}}\left\{ d_Y^{2}(Y_i,\hat{\mu}_{F,m})- d_Y^{2}(Y_i,{\mu}_{F,m})\right\rbrace \Rightarrow -\frac{1}{2}Z_m^\top\Lambda_m^{-1}Z_m,
	\]
	where $Z_m\sim\mathcal N(\mathbf 0,C_m)$, $\Lambda_m=H F_m(\mu_{F,m})$ and
	$C_m=\mathrm{Var}(\psi_m(Y)\mid X\in\Omega_m)$ with $\psi_m(Y)=2\log_{\mu_{F,m}}(Y)$.
\end{proposition}

To better illustrate the conclusion of Proposition~\ref{pro:CLT},
we next present its specialization to the Euclidean setting,
which recovers the well-known quadratic form limit.

\begin{remark}[Euclidean case of Proposition~\ref{pro:CLT}]
	\label{remark:Proposition-CLT}
	In the special case where $Y_1,\dots,Y_n$ are i.i.d.\ random vectors in $\mathbb R^d$ with finite second moment,
	the Fréchet mean coincides with the usual Euclidean mean.
	Specifically, if $Y \in \mathbb R^d$ and $d_Y(y,\theta)=\|y-\theta\|$, then the Fréchet mean is $\mu_F=\mathbb E Y$ and $\log_{\mu_F}(Y)=Y-\mu_F$. Hence
	\[
	\psi(Y)=2\log_{\mu_F}(Y)=2(Y-\mu_F),\quad
	C=\mathrm{Var}(\psi(Y))=4\Sigma,\quad \Sigma:=\mathrm{Var}(Y),
	\]
	and the Hessian of the Fréchet function is $\,\Lambda=HF(\mu_F)=2I_d$ so that $\Lambda^{-1}=\frac12 I_d$.
	Therefore our limit reduces to
	\[
	-\frac12 Z^\top\Lambda^{-1}Z
	= -\frac12 Z^\top\left(\frac12 I_d\right) Z
	= -\frac14 Z^\top Z
	\stackrel{d}{=}
	-\xi^\top \Sigma \xi,
	\]
	where $Z\sim\mathcal N(\mathbf 0,4\Sigma)$ and $\xi\sim\mathcal N(\mathbf 0,I_d)$.
	This coincides with the classical identity
	\[
	\sum_{i=1}^n\left\{\|Y_i-\hat\mu_F\|^2-\|Y_i-\mu_F\|^2\right\}
	= -n\|\bar Y-\mu_F\|^2
	\Rightarrow
	-\xi^\top \Sigma\xi,
	\]
	since $\sqrt n(\bar Y-\mu_F)\Rightarrow \mathcal N(\mathbf 0,\Sigma)$.
	An identical simplification holds cell-wise: for each $m$,
	$\psi_m(Y)=2(Y-\mu_{F,m})$, $C_m=4\Sigma_m$, $\Lambda_m=2I_d$, and the limit becomes
	$-\xi_m^\top \Sigma_m\,\xi_m$ with $\xi_m\sim\mathcal N(\mathbf 0,I_d)$.
\end{remark}

Under $H_0$, the cellwise Fr\'echet means coincide with the global mean, so the fixed-partition null distribution can be expressed in terms of the common center $\mu_F$. The next theorem shows that the rescaled statistic converges to a weighted chi-square law.

\begin{theorem}
	\label{thm:rho-null-fixedM}
	Assume $H_0:\mu_{F,X}=\mu_F$ a.s.\ and
	Assumptions~\ref{assumption:mean-existence}–\ref{assumption:CLT}. For fixed $M$ with $p_m=\mathbb P(X\in\Omega_m)$ and set
	\[
	p=(\sqrt{p_1},\dots,\sqrt{p_M})^\top,\qquad
	\Sigma_W=\operatorname{diag}(C_1,\dots,C_M),\qquad
	D=\operatorname{diag}(\Lambda_1^{-1},\dots,\Lambda_M^{-1}),
	\]
	where $C_m=\mathrm{Var}(\psi(Y)\mid X\in\Omega_m)$ and
	$\Lambda_m=H F_m(\mu_F)$, $\Lambda=HF(\mu_F)$.
	Let
	\[
	B := \Sigma_W^{1/2}\left(D-(p p^\top)\otimes\Lambda^{-1}\right)\Sigma_W^{1/2}
	\in \mathbb R^{Md\times Md}.
	\]
	Here $d:=\dim(T_{\mu_F}\mathcal Y)$ is the intrinsic tangent-space dimension of $\mathcal Y$ at $\mu_F$.
	Then
	\[
	n\rho_n^{M}\Rightarrow \frac{1}{2V_F}\sum_{\ell=1}^{Md}\gamma_\ell Z_\ell^2,
	\]
	where $Z_\ell\stackrel{\mathrm{i.i.d.}}{\sim}\mathcal N(0,1)$ and
	$\{\gamma_\ell\}_{\ell=1}^{Md}$ are the eigenvalues of $B$ (counted with multiplicity).
\end{theorem}

Theorem~\ref{thm:rho-null-fixedM} gives the general null limit in the fixed-partition regime. The following remark shows that, in a simple Gaussian setting, this limit reduces to the familiar ANOVA law.

\begin{remark}[A special case of Theorem~\ref{thm:rho-null-fixedM}]
	Assume $X$ and $Y$ are independent, with $Y_i\stackrel{\mathrm{i.i.d.}}{\sim}\mathcal N(\mu,\sigma^2)$.
	Then the $Y_i$'s have the same distribution across all cells, so $H_0$ holds automatically.
	Define
	\[
	\bar Y_m=\frac1{n_m}\sum_{i:\,X_i\in\Omega_m}Y_i,\qquad
	\bar Y=\frac1n\sum_{i=1}^n Y_i,
	\]
	and
	\[
	S_B=\sum_{m=1}^M n_m(\bar Y_m-\bar Y)^2,\quad
	S_W=\sum_{m=1}^M\sum_{i:\,X_i\in\Omega_m}(Y_i-\bar Y_m)^2,\quad
	S_T=S_B+S_W.
	\]
	So $\frac{S_B}{\sigma^2}\sim \chi^2_{M-1}$, Moreover, since $\hat V_F=S_T/n\xrightarrow{P}\sigma^2$,
	Slutsky’s theorem yields
	\begin{align*}
		n\rho_n^M
		&= \frac{1}{\hat{V}_F}\left[\sum_{i=1}^{n} d_Y^2(Y_i,\hat{\mu}_F) - \sum_{m=1}^{M}\sum_{i=1}^{n}\mathbf 1\{X_i\in\Omega_m\}d_Y^{2}(Y_i,\hat{\mu}_{F,m})  \right]
		= \frac{nS_B}{S_T} \Rightarrow \chi^2_{M-1}.
	\end{align*}
	Thus the classical ANOVA limit law is recovered as a direct special case of Theorem~\ref{thm:rho-null-fixedM}.
\end{remark}

We now turn to the diverging-partition regime $M_n\to\infty$. In this case, the limiting behavior depends on whether the cumulative variance term diverges or remains bounded. The next two theorems treat these two possibilities.

\begin{theorem}
	\label{thm:rho-null-infinite}
	Under $H_0: \mu_{F,X}=\mu_F$ a.s. and Assumptions~\ref{assumption:mean-existence}–\ref{assumption:CLT},
	suppose in addition that
	\[
	\sum_{m=1}^{M_n}\mathrm{tr}\left( (\Lambda_m^{-1}C_m)^2\right) \to \infty ,\qquad
    M_n \asymp {n}^{a}\,(0<a<\frac{1}{3}).
	\]
	Then
	\[
	\frac{n\rho_n^{M_n}-\frac{1}{2\hat V_F}\hat\mu_n}{\frac{1}{2\hat V_F}\hat\sigma_n}
	\Rightarrow \mathcal N(0,1),
	\]
	where
	\[
	\hat\mu_n := \sum_{m=1}^{M_n}\mathrm{tr}\left(\hat\Lambda_m^{-1}\hat C_m\right),\qquad
	\hat\sigma_n^2 := 2\sum_{m=1}^{M_n}\mathrm{tr}\left((\hat\Lambda_m^{-1}\hat C_m)^2\right),
	\]
	with
	\[
	\hat C_m := \frac{1}{n_m}\sum_{i:X_i\in\Omega_m}
	\psi(Y_i)\psi(Y_i)^\top,\qquad
	\hat\Lambda_m := \frac{1}{n_m}\sum_{i:X_i\in\Omega_m}
	\nabla^2 d_Y^2(Y_i,\hat\mu_{F,m}).
	\]
\end{theorem}

Theorem~\ref{thm:rho-null-infinite} covers the regime in which the cumulative variance term diverges, leading to an asymptotically normal limit after suitable centering and scaling. We now turn to the complementary bounded-variance regime, where the limit becomes a second-order Gaussian chaos rather than a Gaussian distribution.

\begin{theorem}
	\label{thm:rho-null-infinite-2}
	Under $H_0: \mu_{F,X}=\mu_F$ a.s. and Assumptions~\ref{assumption:mean-existence}–\ref{assumption:CLT},
	suppose in addition that
	\[
	\lim\limits_{n \to \infty}\sum_{m=1}^{M_n}\mathrm{tr}\left( (\Lambda_m^{-1}C_m)^2\right) < \infty ,\qquad
    M_n \asymp {n}^{a}\,(0<a<\frac{1}{3}).
	\]
	Define
	\begin{align*}
		X&:=\sum_{m\ge1}\left(Z_m^\top\Lambda_m^{-1}Z_m-\mathrm{tr}(\Lambda_m^{-1}C_m)+ p_m\mathrm{tr}(\Lambda^{-1}C_m)\right)
		- \left(\sum_{m\ge1}\sqrt{p_m} Z_m\right)^\top\Lambda^{-1}\left(\sum_{m\ge1}\sqrt{p_m} Z_m\right),
	\end{align*}
	where $Z_m\sim \mathcal N(0,C_m)$ are independent Gaussian vectors.
	Define also the centering terms
	\[
	\hat\mu_n^\star:=\sum_{m=1}^{M_n}\left[
	\mathrm{tr}(\hat\Lambda_m^{-1}\hat C_m)
	-\frac{n_m}{n}\mathrm{tr}(\hat\Lambda^{-1}\hat C_m)
	\right],
	\]
	where $\widehat\Lambda_m,\widehat C_m$ are defined as in Theorem~\ref{thm:rho-null-infinite}, and $\hat \Lambda :=\frac{1}{n}\sum_{i=1}^n
	\nabla^2 d_Y^2(Y_i,\hat\mu_{F})$, then
	\[
	n\rho_n^{M_n} -  \frac{1}{2\hat V_F}\hat\mu_n^\star \Rightarrow \frac{1}{2V_F}\sum_{\ell\ge1}\gamma_\ell(\xi_\ell^2-1).
	\]
	where $\{\xi_\ell\}_{\ell\ge1}$ are i.i.d.\ $\mathcal N(0,1)$ and $\{\gamma_\ell\}_{\ell\ge1}$ are real coefficients satisfying $\sum_{\ell\ge1}\gamma_\ell^2= \frac{1}{2}\mathbb{E}(X^2)=\sum_{m \geq 1} \left(  \mathrm{tr}\left(\Lambda_m^{-1}C_m \right)^2\right)  + \mathrm{tr}\left( (\Lambda^{-1}C)^2  \right) <\infty$. Moreover, the numbers $\gamma_\ell$ are exactly the non-zero eigenvalues (counted with multiplicities) of the corresponding compact symmetric operator
	\[
	\widetilde{B}:  \xi \to \frac{1}{2}\pi_1(X\xi) \,\, \text{on}\,\, H', \quad \xi \in H',
	\]
	where $H'$ is a Gaussian Hilbert space. Here $\pi_1$ denotes the orthogonal projection onto the first-order component of the Gaussian Hilbert space, i.e., the linear span of the underlying Gaussian random variables. In particular, $\widetilde{B} \xi_j= \gamma_\ell \xi_j$.
\end{theorem}

\begin{remark}
	In Theorem~\ref{thm:rho-null-infinite-2}, the notation
	$\sum_{m\geq 1}$ should be interpreted as
	$\lim\limits_{n \to \infty}\sum_{m=1}^{M_n}$,
	that is, for each finite $n$ the index $m$ is always restricted to $m \leq M_n$.
	This convention ensures that the infinite sums in the limit are understood as limits of the corresponding finite sums.
\end{remark}

These results characterize the null behavior of $\rho_n^{M_n}$ on finite-dimensional Riemannian manifolds in both fixed and diverging partition regimes. We next treat one-dimensional Wasserstein responses, where the argument proceeds through the $L^2$ quantile embedding rather than local manifold coordinates.

\subsubsection{One-dimensional Wasserstein responses: null asymptotics via the $L^2$ embedding}
\label{subsubsec:null-w2}

The results above rely on manifold-based arguments.
For one-dimensional Wasserstein responses, the sample space $\mathcal W_2(\mathbb R)$ is not a finite-dimensional Riemannian manifold, so those arguments are not directly applicable.
Instead, we exploit the quantile representation of the $2$-Wasserstein distance, which yields an isometric embedding into the Hilbert space $H=L^2(0,1)$.
This allows us to re-establish the same three null limits (fixed $M$, $\sigma_n^2\to\infty$ CLT, and $\sup_n\sigma_n^2<\infty$ second-chaos limit) by Hilbert-space arguments.

In this subsubsection we focus on distributional responses that take values in a bounded subset of the Wasserstein space $\mathcal W_2(\mathbb R)$ endowed with the $2$-Wasserstein distance.
For distributions $F$ and $G$, the squared distance is
\[
d_W^2(F,G)=\int_0^1\big(F^{-1}(s)-G^{-1}(s)\big)^2\,ds,
\]
where $F^{-1}$ and $G^{-1}$ are the corresponding quantile functions.
For $F\in\mathcal W_2(\mathbb R)$, let $Q_F:=F^{-1}$. Then the map $F\mapsto Q_F$ is an isometric embedding into
$H:=L^2(0,1)$, in the sense that $d_W(F,G)=\|Q_F-Q_G\|_H$.

\begin{theorem}
	\label{thm:rho-null-fixedM-wass}
	Assume $M_n\equiv M$ is fixed and the responses satisfy $Y\in\mathcal W_2(\mathbb R)$. Under $H_0:\mu_{F,X}=\mu_F$ a.s.\ and Assumptions~\ref{assumption:mean-existence}--\ref{assumption:cell-prob}, we have
	\[
	n\rho_n^{M_n} \Rightarrow \frac{1}{V_F}\sum_{\ell\ge 1}\gamma_\ell \zeta_\ell^2,
	\]
	where $\zeta_\ell\stackrel{i.i.d.}{\sim}N(0,1)$ and $\{\gamma_\ell\}_{\ell\ge 1}$ are the eigenvalues (counted with multiplicity) of the
	self-adjoint, nonnegative trace-class operator
	\[
	\mathcal B:=\Sigma_W^{1/2}\,\mathcal P\,\Sigma_W^{1/2}\quad\text{on }H^M,
	\qquad
	\Sigma_W:=\mathrm{diag}(\Sigma_1,\dots,\Sigma_M),
	\qquad
	\mathcal P:=I_{H^M}-(pp^\top)\otimes I_H,
	\]
	with $H:=L^2(0,1)$, $p_m:=\mathbb P(X\in\Omega_m)>0$, $p:=(\sqrt{p_1},\dots,\sqrt{p_M})\in\mathbb R^M$, and
	\[
	\Sigma_m:=\mathrm{CovOp}(Q\mid X\in\Omega_m),
	\qquad
	Q:=F_Y^{-1}\in H.
	\]
\end{theorem}

\begin{remark}
	\label{remark: wass-trace-class}
	Since $\mathbb E\|Q\|_H^2<\infty$, each conditional covariance operator
	$\Sigma_m=\mathrm{CovOp}(Q\mid X\in\Omega_m)$ is self-adjoint, nonnegative and trace-class on $H$; hence
	$\mathrm{tr}(\Sigma_m)$ and $\mathrm{tr}(\Sigma_m^2)$ are well-defined and finite.
\end{remark}

Theorem~\ref{thm:rho-null-fixedM-wass} gives the null limit when the number of cells is fixed.
We next let the partition $M_n\to\infty$. In this regime, the asymptotic behavior is governed by the aggregate size of the cellwise covariance operators, summarized by
\(
\sigma_n^2=2\sum_{m=1}^{M_n}\mathrm{tr}(\Sigma_m^2).
\)
Depending on whether $\sigma_n^2$ diverges or stays bounded, we obtain two different null limits.

\begin{theorem}
	\label{thm:rho-null-infinite-wass}
	Assume $Y\in\mathcal W_2(\mathbb R)$ and let $Q:=F_Y^{-1}\in H:=L^2(0,1)$.
	Under $H_0:\mu_{F,X}=\mu_F$ a.s.\ and Assumptions~\ref{assumption:mean-existence}--\ref{assumption:cell-prob}, suppose in addition that
	\[\sigma_n^2:=2\sum_{m=1}^{M_n}\mathrm{tr}\!\left(\Sigma_m^2\right)\to\infty,\qquad
    M_n \asymp {n}^{a}\,(0<a<\frac{1}{3}).
	\]
	where $\Sigma_m:=\mathrm{CovOp}(Q\mid X\in\Omega_m)$.
	Define
	\[
	\hat\mu_n:=\sum_{m=1}^{M_n}\mathrm{tr}(\hat\Sigma_m),
	\qquad
	\hat\sigma_n^2:=2\sum_{m=1}^{M_n}\mathrm{tr}\!\left(\hat\Sigma_m^2\right),
	\]
	with
	\[
	\hat\Sigma_m:=\frac{1}{n_m}\sum_{i:X_i\in\Omega_m}(Q_i-\bar Q_m)\otimes(Q_i-\bar Q_m),
	\qquad
	\bar Q_m:=\frac{1}{n_m}\sum_{i:X_i\in\Omega_m}Q_i,
	\]
	Then
	\[
	\frac{n\rho_n^{M_n}-\hat\mu_n/\hat V_F}{\hat\sigma_n/\hat V_F}\Rightarrow \mathcal N(0,1).
	\]
\end{theorem}

When $\sigma_n^2\to\infty$, the studentized statistic satisfies a central limit theorem as stated in Theorem~\ref{thm:rho-null-infinite-wass}.
In contrast, if $\sigma_n^2$ remains bounded, the limiting distribution is no longer Gaussian.
The next theorem characterizes this bounded-variance regime and shows that the limit can be represented as a second-order Gaussian chaos with an associated spectral decomposition.

\begin{theorem}
	\label{thm:rho-null-infinite-wass-2}
	Assume $Y\in\mathcal W_2(\mathbb R)$ and let $Q:=F_Y^{-1}\in H:=L^2(0,1)$.
	Under $H_0:\mu_{F,X}=\mu_F$ a.s.\ and Assumptions~\ref{assumption:mean-existence}--\ref{assumption:cell-prob},
	suppose in addition that
	\[
	\sup_n \sigma_n^2<\infty,\qquad
    M_n \asymp {n}^{a}\,(0<a<\frac{1}{3}).
	\]
	where $\Sigma_m:=\mathrm{CovOp}(Q\mid X\in\Omega_m)$, $\sigma_n^2:=2\sum_{m=1}^{M_n}\mathrm{tr}(\Sigma_m^2)$.
	Let $V_F:=\mathbb E\|Q-\mathbb EQ\|_H^2\in(0,\infty)$, $\Sigma:=\mathrm{CovOp}(Q)=\sum_{m\ge1}p_m\Sigma_m $.
	Define $\hat V_F$, $\hat\Sigma_m$, and $\hat\mu_n:=\sum_{m=1}^{M_n}\mathrm{tr}(\hat\Sigma_m)-\hat V_F$
	as in Theorem~\ref{thm:rho-null-infinite-wass}.
	Let $\{G_m\}_{m\ge1}$ be independent Gaussian elements $G_m\sim N_H(0,\Sigma_m)$, and define
	\[
	X:=\sum_{m\ge1}\Big(\|G_m\|_H^2-\mathrm{tr}(\Sigma_m)\Big)
	-\Big(\Big\|\sum_{m\ge1}\sqrt{p_m}\,G_m\Big\|_H^2-\mathrm{tr}(\Sigma)\Big).
	\]
	Then there exist real coefficients $\{\gamma_\ell\}_{\ell\ge1}$ with
	$\sum_{\ell\ge1}\gamma_\ell^2<\infty$ such that
	\[
	n\rho_n^{M_n}-\frac{\hat\mu_n}{\hat V_F}\ \Rightarrow\frac{1}{V_F}\sum_{\ell\ge1}\gamma_\ell(\zeta_\ell^2-1),
	\quad
	\zeta_\ell\stackrel{i.i.d.}{\sim}N(0,1).
	\]
\end{theorem}

In summary, we have established the null limits of $\rho_n^{M_n}$ under $H_0$ both on finite-dimensional Riemannian manifolds and for one-dimensional Wasserstein responses (via the $L^2$ quantile embedding), thereby providing a unified basis for inference with fixed or growing partitions.

\subsection{Fixed-partition wild bootstrap}\label{subsec:wild-bootstrap}
We now revisit the null hypothesis~\eqref{eq:null} from a practical perspective. 
Inspiringly, results in Theorems~\ref{thm:rho-null-fixedM} and \ref{thm:rho-null-fixedM-wass} motivate
a wild bootstrap procedure for FCC to test~\eqref{eq:null}, based on a restricted resampling scheme for the linearized null score process. 
We defer the detailed implementation and algorithm to Appendix~\ref{appendix:wild-bootstrap-algorithm} and explain only the core idea here.

Briefly, we impose the null restriction before resampling. Conditional on the fixed partition, we center the response at the global Fr\'echet fit rather than at cell-specific fits, and multiply the resulting centered residuals or tangent scores by i.i.d. multipliers of zero mean and unit variance. 
For each bootstrap replicate, we recompute the between-cell component while keeping the original partition and scaling fixed. 
The bootstrap therefore perturbs only the linearized null fluctuation appearing in the fixed-$M$ limits, while preserving the empirical cell sizes and allowing the conditional variability of the response to vary across cells. 
The theoretical justification is given by the following theorem.
\begin{theorem}\label{thm:bootstrap-validity-fixedM-unified}
Suppose $M$ is fixed and $H_0:\mu_{F,X}=\mu_F$ almost surely. Let $T_{\mathrm{obs}}$ and $T_b^*$ be defined in Algorithm~\ref{alg-wild} of Appendix~\ref{appendix:wild-bootstrap-algorithm}. Assume that the multipliers $\{\xi_i\}_{i=1}^n$ are i.i.d., independent of the data, and satisfy $\mathbb E(\xi_i)=0$, $\mathbb E(\xi_i^2)=1$, and $\mathbb E|\xi_i|^{2+\eta}<\infty$ for some $\eta>0$. Assume either that $\mathcal Y$ is a finite-dimensional Riemannian manifold and Assumptions~\ref{assumption:mean-existence}--\ref{assumption:CLT} hold, or that $Y\in W_2(\mathbb R)$, the response space is contained in a bounded subset of $W_2(\mathbb R)$, and Assumptions~\ref{assumption:mean-existence}--\ref{assumption:cell-prob} hold. Then
\[
\sup_{t\in\mathbb R}\left|\mathbb P^*(T_b^*\le t)-\mathbb P(T_{\mathrm{obs}}\le t)\right|\to 0
\qquad\text{in probability},
\]
and
$
T_{\mathrm{obs}}-n\rho_n^M=o_P(1).
$
Hence the observed statistic $T_{\mathrm{obs}}$ in Algorithm~\ref{alg-wild} can be asymptotically replaced by $n\rho_n^M$. Here \(\mathbb P^*(\cdot):=\mathbb P(\cdot\mid \{(X_i,Y_i)\}_{i=1}^n)\). In particular, the fixed-$M$ wild bootstrap consistently calibrates the null law of \(T_{\mathrm{obs}}\).
\end{theorem}

In summary, our wild bootstrap procedure is tailored to the FCC null~\eqref{eq:null}. 
It recentres the response at the global Fr\'echet fit and does not permute the pairs $(X_i,Y_i)$, thus preserves the fixed partition, the empirical cell sizes, and possible cellwise heteroskedasticity under $H_0$. 
It provides a practical testing framework for directional Fr\'echet-mean dependence across distinct metric spaces.

\subsection{FCC under specific stochastic models}\label{subsec:model-specific}
Beyond asymptotic inference, it is useful to understand how FCC behaves under structured stochastic models. Although FCC need not be monotone in a generic noise parameter for arbitrary models, the next two examples show that, in representative Wasserstein and SPD settings, FCC decreases as the noise level $\sigma$ increases.

\subsubsection{Wasserstein Data}\label{subsubsec:wasserstein data}
We consider response distributions in the Wasserstein space $\mathcal{W}_2(\mathbb{R})$, equipped with the squared $2$-Wasserstein distance
$$
d_W^2(F,G) = \int_0^1 \left(F^{-1}(s) - G^{-1}(s)\right)^2 \, ds,
$$
where $F^{-1}$ and $G^{-1}$ are the quantile functions. 
We adopt two models from \citet{bhattacharjee2023single}. They are intended as illustrative examples rather than universal generative mechanisms, but they provide tractable settings in which the effect of the noise parameter can be analyzed explicitly.

\textbf{Setting I:}
\begin{itemize}
	\item Quantile function: $Q(Y)(\cdot) = \mu + t\Phi^{-1}(\cdot),$
	\item Location parameter: $\mu \sim \mathcal{N}(\zeta(X), \sigma^2),$
	\item Scale parameter: $t \sim \text{Exp}\left(\frac{X}{1 + \exp(X)}\right).$
\end{itemize}

\textbf{Setting II:}
\begin{itemize}
	\item Quantile process: $Q(Y)(\cdot) = T_k(\mu + t\Phi^{-1}(\cdot)),$
	\item Location parameter: $\mu \sim \mathcal{N}(\zeta(X), \sigma^2),\ t = 0.1,$
	\item Transport maps: $T_k(a) = a - \frac{\sin(ka)}{|a|},\ k \in \{\pm1,\pm2,\pm3\}.$
\end{itemize}
For Setting II, the transport map is drawn uniformly from the family $\{T_k:k\in\{\pm1,\pm2,\pm3\}\}$ and then applied to the base quantile function. The parameters $\mu$ and $t$ are sampled independently.

\begin{proposition}\label{pro:rho wasserstein data}
	In the contexts of both Setting I and Setting II, the correlation \(\rho\) decreases as \(\sigma\) increases.
\end{proposition}

\subsubsection{SPD data under the log-Euclidean metric}\label{subsubsec:log_euclidean}
Let $\mathcal{P}_m$ denote the manifold of $m \times m$ symmetric positive-definite (SPD) matrices equipped with the \textit{log-Euclidean metric}.
To illustrate the behavior of $\rho$ in this setting, we adopt two simulation models considered in \citet{zhang2024dimension}.
These models are not intended as general generative mechanisms for SPD data, but rather as stylized examples that have been used in prior simulation studies.

Specifically, $\log(Y)$ is generated according to $N_{dd}(\log(D(X)), \sigma^2)$, that is,
$\log(Y) = \sigma Z + \log(D(X))$, where $\log(\cdot)$ denotes the matrix logarithm, $Z$ is a symmetric matrix with independent $\mathcal{N}(0,1)$ entries on the diagonal and $\mathcal{N}(0,\frac{1}{2})$ on the off-diagonal elements. The matrix $D(X)$ is specified according to the following models:

\noindent \textbf{Setting I: }
$$
D(X) =
\begin{pmatrix}
	1 & \rho(X) \\
	\rho(X) & 1
\end{pmatrix}, \quad
\rho(X) =  \frac{\exp(X) - 1}{\exp(X) + 1}.
$$

\noindent \textbf{Setting II: }
$$
D(X) =
\begin{pmatrix}
	1 & \rho_1(X) &\rho_2(X) \\
	\rho_1(X) & 1 & \rho_1(X) \\
	\rho_2(X) & \rho_1(X) & 1
\end{pmatrix}, \quad
		\rho_1(X) =  0.4 \frac{\exp(X) - 1}{\exp(X) + 1}, \quad \rho_2(X) =  0.4 \sin(X).
$$

\begin{proposition}\label{pro:rho spd data}
	In the context of data on the manifold of symmetric positive-definite (SPD) matrices under the log-Euclidean metric, for both Setting I and Setting II, the correlation \(\rho\) decreases as \(\sigma\) increases.
\end{proposition}

Taken together, Propositions~\ref{pro:rho wasserstein data} and \ref{pro:rho spd data} show that, in these structured Wasserstein and SPD models, increasing the noise level attenuates FCC. These examples do not imply a universal monotonicity law, but they do support the interpretation of FCC as a meaningful explained-variation coefficient in representative non-Euclidean settings.

This completes the theoretical development of FCC: we have established its basic bounds and boundary cases, proved consistency of the partition estimator, derived null asymptotic distributions in the main geometric regimes, and illustrated its behavior under representative stochastic models.

\section{Simulation}\label{sec:simulation}
We evaluate finite-sample rejection behavior in one Euclidean benchmark and four non-Euclidean settings. For Settings I--IV, the reported rejection rates correspond to empirical power against structured alternatives with nontrivial Fr\'echet-mean signal. 
Setting V is a dependent but under the null~\eqref{eq:null}, designed to illustrate the deliberate scope of FCC relative to omnibus independence procedures.
We briefly summarize the experimental setup here and defer the data-generating models and parameter configurations for the reported figures to Appendix~\ref{sec:sim-details}.

Setting I is a sparse Euclidean vector model in which only the first coordinate pair $(X_{i1},Y_{i1})$ carries a nonlinear signal, while the remaining coordinates are Gaussian noise. 
Setting II is a spherical model on $\mathbb{S}^2$, where a latent phase determines $X$ and $Y$ through different transformations. Setting III is a Wasserstein model in which the predictor is a location family and the response carries a periodic location signal. 
Setting IV is an SPD model under the Log--Cholesky metric with a folded conditional Fr\'echet mean structure, so it serves as a non-Euclidean power benchmark favorable to the FCC target.
Setting V is a Wishart SPD benchmark with sign-symmetric shape modulation, designed so that the directional Fr\'echet-mean signal is largely canceled while the conditional covariance still changes with $X$.

\begin{figure}[!ht]
	\centering
	\begin{minipage}[t]{0.5\linewidth}
		\centering
		\includegraphics[width=\linewidth]{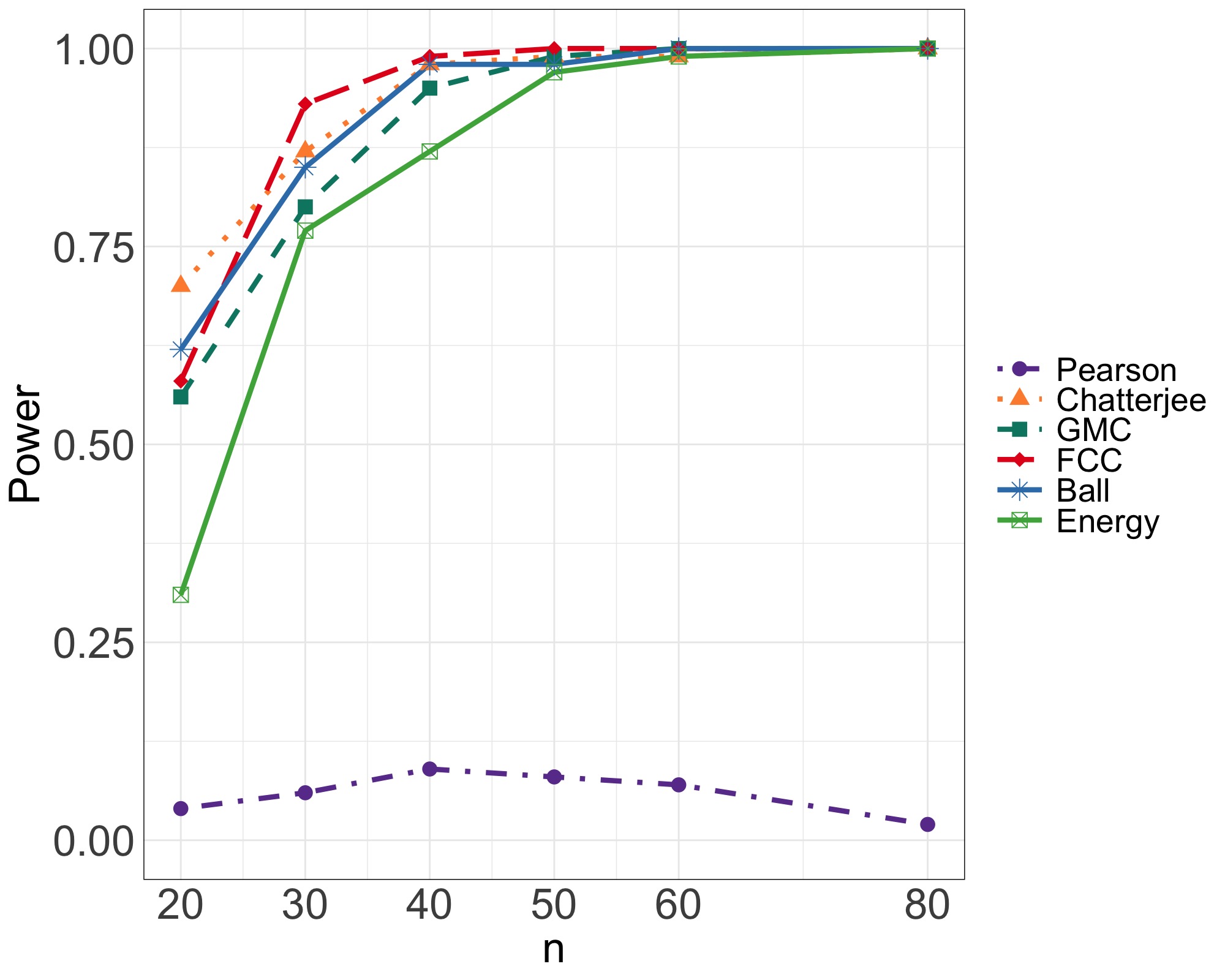}
		{\small (a) $\mathbb{R}$.}
	\end{minipage}
	\begin{minipage}[t]{0.48\linewidth}
		\centering
		\includegraphics[width=\linewidth]{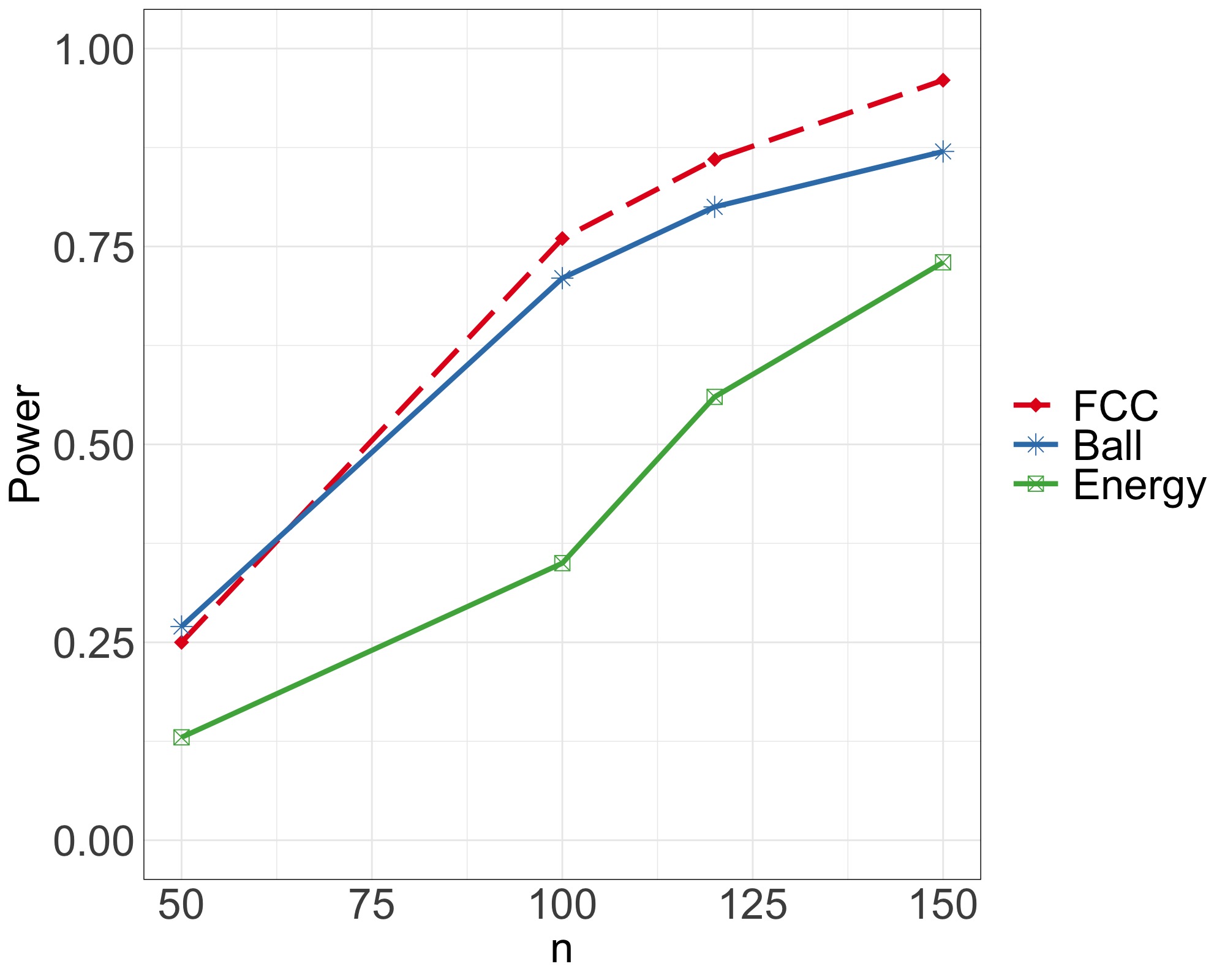}
		{\small (b) $\mathbb{R}^3$.}
	\end{minipage}
	\caption{Empirical rejection rate at the nominal level $0.05$ against sample size $n$ for Setting I. Panel (a) shows the scalar benchmark, in which all six methods are applied to the first coordinate pair $(X_{i1}, Y_{i1})$. Panel (b) reports the results for the full sparse vector setting.}
	\label{fig:sim1_scalar}
\end{figure}

In each panel we generate i.i.d. pairs $\{(X_i,Y_i)\}_{i=1}^n$ and report empirical rejection rates at nominal level $\alpha=0.05$ as functions of the sample size $n$. For the scalar benchmark in Figure~\ref{fig:sim1_scalar}(a), we compare Pearson correlation, Chatterjee's rank-based coefficient \citep{chatterjee2021new}, generalized measures of correlation (GMC) \citep{zheng2012generalized}, Ball covariance \citep{pan2020ball}, Energy distance covariance \citep{szekely2007measuring}, and FCC on the same one-dimensional signal; the benchmark procedures are permutation-calibrated, whereas FCC uses the fixed partition wild bootstrap from Appendix~\ref{appendix:wild-bootstrap-algorithm}. 
For the non-Euclidean panels, we compare FCC, Ball, and Energy under the natural metric of each space.

\begin{figure}[!ht]
  \centering
  \begin{minipage}[t]{0.46\linewidth}
    \centering
    \includegraphics[width=\linewidth]{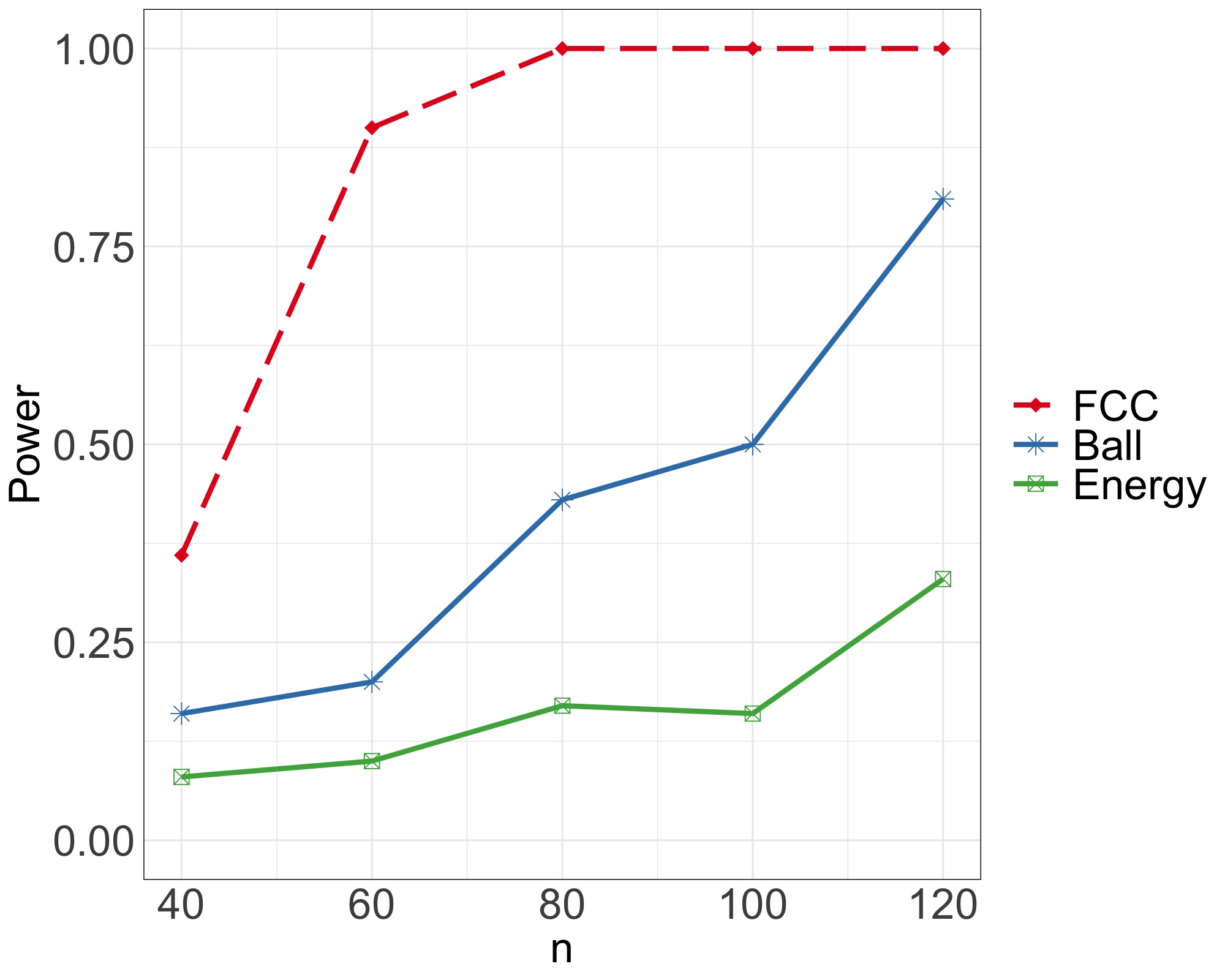}
    {\small (a) $\mathbb{S}^2$}
  \end{minipage}\hfill
  \begin{minipage}[t]{0.46\linewidth}
    \centering
    \includegraphics[width=\linewidth]{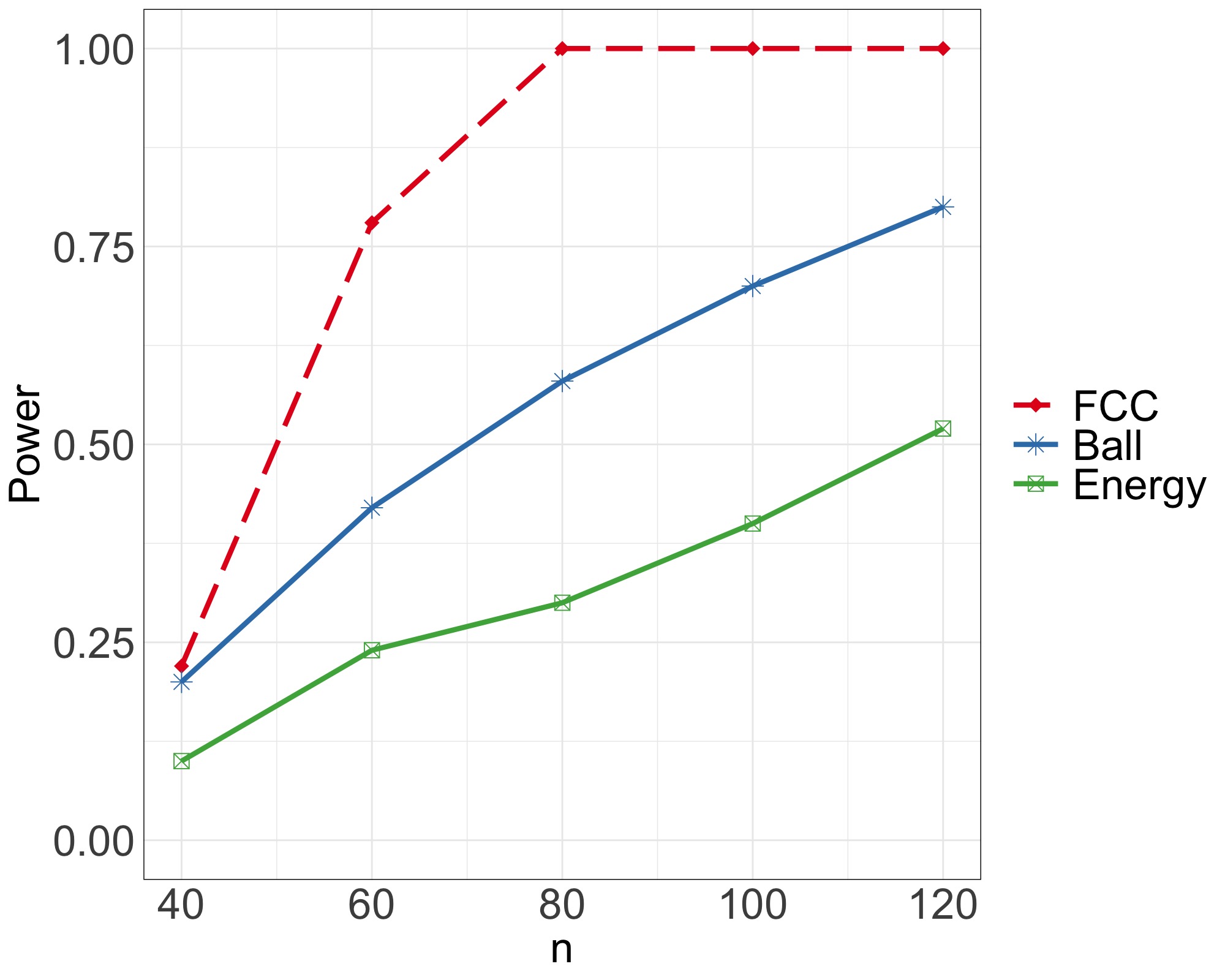}
    {\small (b) $\mathcal{W}_2$}
  \end{minipage}
  \medskip
  \begin{minipage}[t]{0.46\linewidth}
    \centering
    \includegraphics[width=\linewidth]{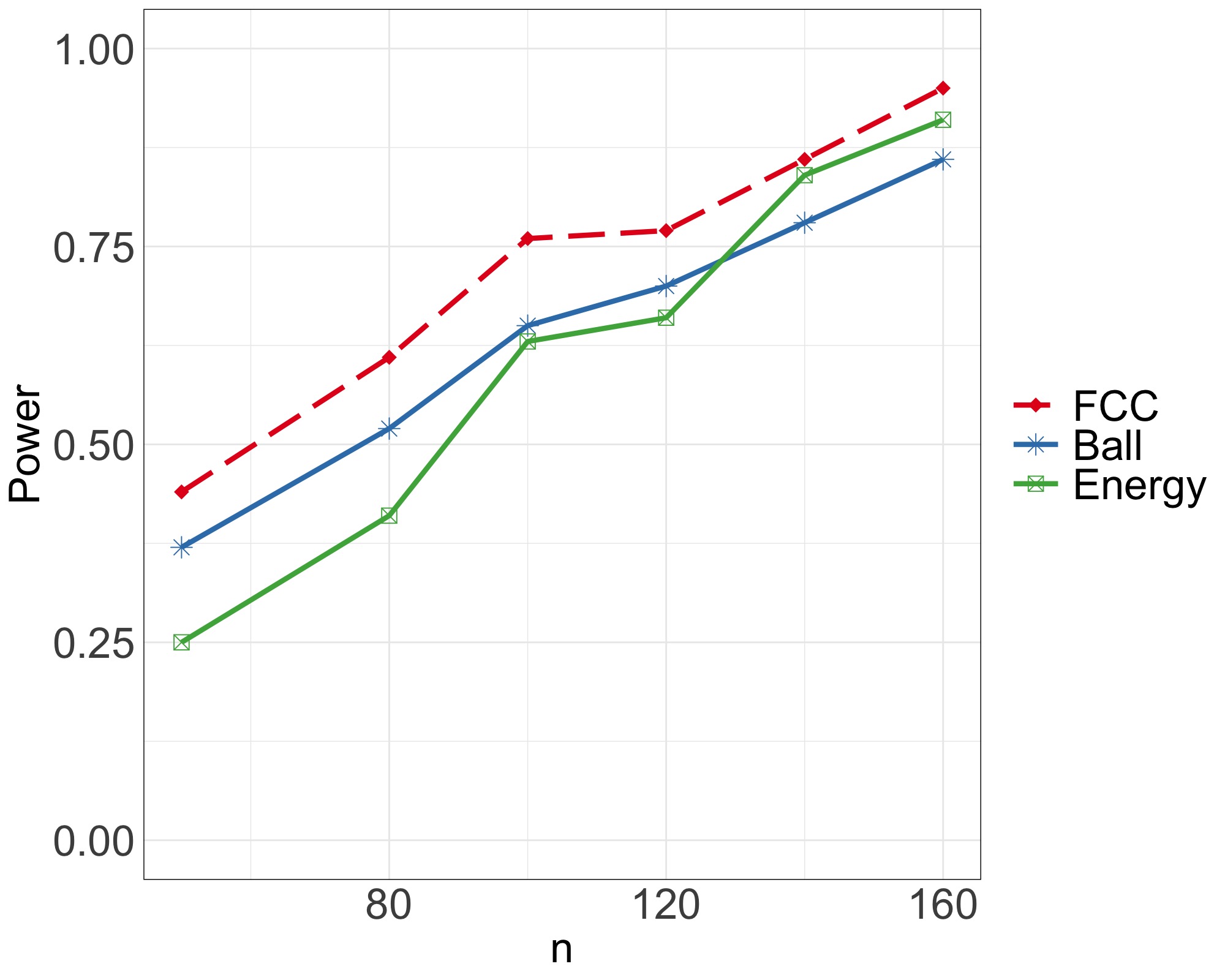}
    {\small (c) $\mathcal{S}_4^+$}
  \end{minipage}\hfill
  \begin{minipage}[t]{0.46\linewidth}
    \centering
    \includegraphics[width=\linewidth]{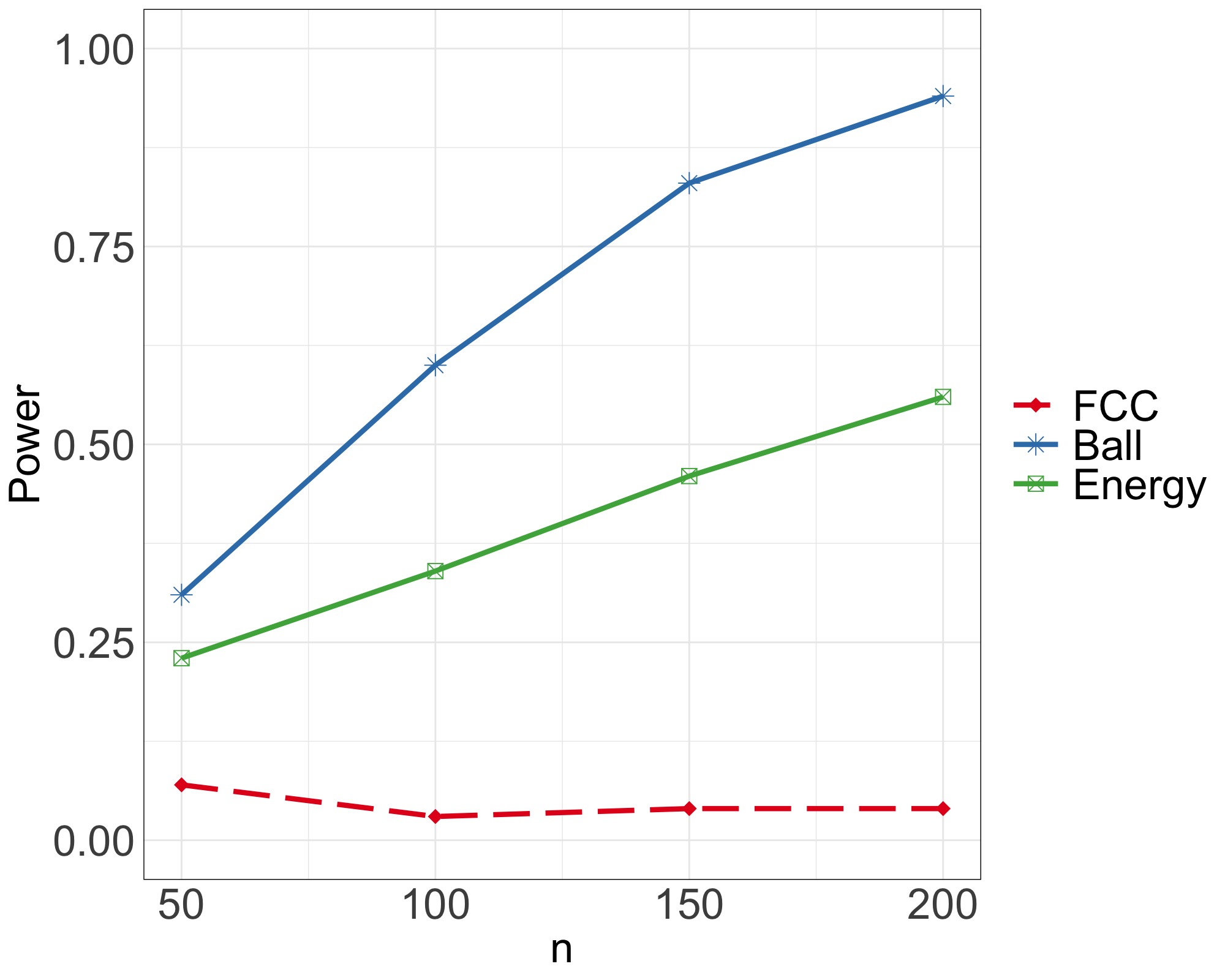}
    {\small (d) $\mathcal{S}_4^+$ under the null~\eqref{eq:null}}
  \end{minipage}
  \caption{Empirical rejection rates at the nominal level $0.05$ against sample size $n$ for the non-Euclidean settings II--V. Panels (a), (b), and (c) are power comparisons under alternatives with nonzero FCC, whereas panel (d) is a dependent but without a conditional Fr\'echet mean signal test. Exact panel-specific numerical settings are given in Appendix~\ref{sec:sim-details}.}
  \label{fig:sim1}
\end{figure}

Figure~\ref{fig:sim1_scalar} (a) isolates the scalar nonlinear signal and shows that Chatterjee, GMC, Ball, FCC, and Energy all attain high power rapidly, whereas Pearson remains close to the nominal level because the dependence is strongly nonlinear. Figure~\ref{fig:sim1_scalar} (b) then embeds the same qualitative signal in a sparse vector problem and shows that all three metric-space procedures improve with $n$, with energy increasing more slowly in this benchmark.

For the non-Euclidean settings II--V, Figure~\ref{fig:sim1} compares FCC, Ball, and Energy using the metric natural to each space. Panels (a)--(c) are results for mean-structured models, and all three methods gain power with $n$, with FCC generally the most competitive there.
Panel (d) serves a different purpose: although the model is under the null hypothesis~\eqref{eq:null}, it is still dependent. Accordingly, FCC is expected to stay near its nominal level, whereas Ball and Energy can still reject because they target generic dependence rather than the conditional Fr\'echet-mean effects. 
Taken together, these simulations highlight both the strengths of the FCC and its focuses on Fr\'echet-mean explained variation rather than general dependence.

\section{Conclusion}\label{sec:conclusion}
We introduces the Fr\'echet correlation coefficient (FCC) as a directional, model-free measure of explained variation for responses and predictors taking values in general metric spaces.
FCC is designed to extend the classical notion of $R^2$ to the non-Euclidean setting.
At the population level, FCC has a clear interpretation: it lies in $[0,1]$, equals one if and only if the response is almost surely a measurable function of the predictor, and equals zero when conditioning on the predictor does not change the Fr\'echet mean of the response. Accordingly, FCC serves as a response-specific effect size measure for Fr\'echet mean dependence, rather than a general coefficient for arbitrary forms of dependence.

To make FCC operational, we propose an efficient partition-based estimator. We establish its consistency under mild conditions and derive null asymptotic theory under both fixed-partition and growing-partition regimes. The resulting inference procedure is computationally straightforward and applies uniformly across heterogeneous predictor geometries.

Several directions merit further study. Methodologically, it would be useful to develop data-adaptive partition schemes with stronger finite-sample guarantees and to investigate resampling or calibration schemes that improve small-sample inference.
From a modeling perspective, an important extension is to connect FCC more systematically with Fr\'echet regression and related conditional mean models, particularly in settings where one seeks both an overall effect-size summary and an interpretable regression surface.

\appendix

\section{Implementation Details for the Wild Bootstrap in Section~\ref{subsec:wild-bootstrap}}\label{appendix:wild-bootstrap-algorithm}
This section provides implementation details for the wild bootstrap procedure introduced in Section~\ref{subsec:wild-bootstrap} of the main text. 
Throughout this section, we consider a  fixed partition $\Pi=\{\Omega_m\}_{m=1}^{M}$ and define  $n_m=\sum_{i=1}^n\mathbf 1\{X_i\in\Omega_m\}$ and $\hat p_m=n_m/n$. 
The main steps of the wild bootstrap procedure are summarized in Algorithm~\ref{alg-wild}.

\begin{algorithm}[!ht]
	\caption{Fixed-$M$ wild bootstrap for FCC}\label{alg-wild}
	\DontPrintSemicolon
	\KwIn{Data $\{(X_i,Y_i)\}_{i=1}^n$; metrics $d_X,d_Y$; fixed partition $\{\Omega_m\}_{m=1}^{M}$; bootstrap replicates $B$; multiplier law $\xi_i\sim\mathcal L_\xi$ with $\E\xi_i=0$ and $\E\xi_i^2=1$.}
	\KwOut{Bootstrap p-value $\widehat p$.}

	\BlankLine
	\textbf{Step 1: Embed and center}\;
    Construct embedded responses $Z_i$ (Table~\ref{table:alg});
    compute $\bar Z=n^{-1}\sum_{i=1}^nZ_i$ and $Z_i^c=Z_i-\bar Z$;
    compute $n_m$ and $\hat p_m=n_m/n$ for $m=1,\dots,M$.\;

	\BlankLine
    \textbf{Step 2: Preliminary quantities}\;
    Compute the observed FCC estimate $\widehat\rho$, the global scale factor $\hat V_F$, and the normalization matrices/operators $H_m$ and $H$ (Remark~\ref{rem:normalization}).\;

	\BlankLine
    \textbf{Step 3: Observed and bootstrap statistics}\;
    Compute $\bar Z_m$ and $\bar Z$ from $\{Z_i^c\}$ using the original partition, set $
	B_n=\sum_{m=1}^{M}n_m\|H_m^{-1/2}\bar Z_m\|^2-n\|H^{-1/2}\bar Z\|^2$, and
	$T_{\mathrm{obs}}=\frac{B_n}{\hat V_F}$.

	\For{$b=1,\dots,B$}{
		Draw i.i.d. $\xi_1^{(b)},\dots,\xi_n^{(b)}$ from $\mathcal L_\xi$;\;
		$Z_i^{*(b)}=\xi_i^{(b)}Z_i^c,\ i=1,\dots,n$;\;
		Compute $\bar Z_m^{*(b)}$ and $\bar Z^{*(b)}$ from $\{Z_i^{*(b)}\}$ using the original partition, set
        $B_n^{*(b)}=\sum_{m=1}^{M}n_m\|H_m^{-1/2}\bar Z_m^{*(b)}\|^2-n\|H^{-1/2}\bar Z^{*(b)}\|^2$,
        and set $T_b^*=B_n^{*(b)}/\hat V_F$.\;
	}

	\BlankLine
	\textbf{Step 4: Compute p-value}\;
	\[
	\widehat p=\frac{1+\sum_{b=1}^B \mathbf 1\{T_b^*\ge T_{\mathrm{obs}}\}}{B+1}.\;
	\]
\end{algorithm}

\begin{remark}
\label{rem:normalization}
The normalization used in the observed and bootstrap statistics is determined by the local quadratic structure of the response space. When the relevant Hessian matrices/operators are known explicitly, we use their population forms; otherwise, we use plug-in estimates.
\end{remark}

For Step 1 of Algorithm~\ref{alg-wild}, Table~\ref{table:alg} reports the response embeddings used in this paper.
Here, the embedded responses $Z_i$ use the response metric: direct Euclidean coordinates, weighted quantile coordinates, Log--Cholesky coordinates, or the sphere log map at the global intrinsic mean $\hat\mu_F$.
\begin{table}[!ht]
	\caption{Expression of $Z_i$.}
	\begin{tabular}{lll}
		\hline
		Response & Observation & Embedding used in the bootstrap \\ \hline
		Euclidean & $Y_i\in\mathbb R^d$ & $Z_i=Y_i$ \\
		Wasserstein & Quantile $Q_i$ on grid $\{q_\ell\}$ & $Z_i=(\sqrt{w_\ell}\,Q_i(q_\ell))_{\ell}$,  with integration weights $\{w_\ell\}$ \\
		SPD & $Y_i\in\mathcal S_p^+$ & $Z_i=\mathrm{LC}(Y_i)$ (Log--Cholesky coordinate vector) \\
		Sphere & $Y_i\in\mathbb S^{d-1}$ & $Z_i=\log_{\hat\mu_F}(Y_i)$ \\ \hline
	\end{tabular}
	\label{table:alg}
\end{table}
For the Wasserstein embedding, the Euclidean norm of $Z_i-Z_j$ matches the discretized $L^2$ distance between the corresponding quantile functions. For the sphere case, the embedding is local: it linearizes the response around the global intrinsic mean rather than producing an exact global isometry.

In Step~2, the global scale factor is always taken to be the sample Fr\'echet variance
\[
\hat V_F=\frac{1}{n}\sum_{i=1}^n d_Y^2(Y_i,\hat\mu_F).
\]
Let
\[
\bar Z=\frac{1}{n}\sum_{i=1}^n Z_i,\qquad Z_i^c=Z_i-\bar Z,
\]
and set
\[
n_m=\sum_{i=1}^n \mathbf 1\{X_i\in\Omega_m\},\qquad 
\bar Z_m=\frac{1}{n_m}\sum_{i=1}^n \mathbf 1\{X_i\in\Omega_m\}Z_i^c,\quad m=1,\dots,M.
\]
Then the observed statistic is computed by
\[
T_{\mathrm{obs}}=\frac{B_n}{\hat V_F},
\]
where
\[
B_n=\sum_{m=1}^{M}n_m\|H_m^{-1/2}\bar Z_m\|^2-n\|H^{-1/2}\bar Z\|^2.
\]
That is, \(B_n\) is the normalized between-cell quadratic form determined by the cellwise and global normalization matrices/operators \(H_m\) and \(H\). In particular, when the cellwise and global normalization coincide, this reduces to
\[
B_n=\sum_{m=1}^{M}n_m\|H^{-1/2}\bar Z_m\|^2-n\|H^{-1/2}\bar Z\|^2
=\sum_{m=1}^{M} n_m\|H^{-1/2}(\bar Z_m-\bar Z)\|^2.
\]
Moreover, by Theorem~\ref{thm:bootstrap-validity-fixedM-unified}, \(T_{\mathrm{obs}}\) can be asymptotically replaced by \(n\widehat\rho\).

Step 3 uses a restricted multiplier resampling scheme.
For each bootstrap replication,
\[
Z_i^{*(b)}=\xi_i^{(b)}Z_i^c,\qquad i=1,\dots,n,
\]
with i.i.d.\ multipliers $\xi_i^{(b)}$ drawn from $\mathcal L_\xi$. The bootstrap statistic is

\[
T_b^*=\frac{B_n^{*(b)}}{\hat V_F}, \qquad B_n^{*(b)}=\sum_{m=1}^{M}n_m\|H_m^{-1/2}\bar Z_m^{*(b)}\|^2-n\|H^{-1/2}\bar Z^{*(b)}\|^2,
\]
where $\bar Z_m^{*(b)}$ and $\bar Z^{*(b)}$ are the cellwise and global averages of $\{Z_i^{*(b)}\}$. The same partition and the same denominator $\hat V_F$ are kept fixed across all bootstrap draws, so the bootstrap targets the null fluctuation of the between-cell term only.

We make some comments for Algorithm~\ref{alg-wild}.
First, in current experiments,  the predictor partition  $\{\Omega_m\}_{m=1}^{M}$  is constructed once from the auxiliary samples and is held fixed across all bootstrap draws. Thus the procedure is a \emph{fixed-partition}  calibration.
Second, the bootstrap is \emph{restricted to the FCC null}: the response is always centered at the global Fr\'echet fit, rather than at cell-specific fits, before the multipliers are applied. This is what makes the method target $H_0:\mu_{F,X}=\mu_F$ rather than independence.
Third,  the denominator $\hat V_F$ is fixed at its observed value throughout the bootstrap, and only resamples the between-cell fluctuation term. This mirrors the fixed-$M$ asymptotic theory, in which the main null randomness of $n\widehat\rho$ comes from the cellwise score process, while the global Fr\'echet variance is estimated consistently and enters only as a plug-in scale factor.

\begin{remark}
	Our bootstrap procedure is a geometry-aware scheme tailored to FCC to handle the null hypothesis~\eqref{eq:null}. Its contribution is at least twofold. First, it turns the fixed-$M$ asymptotic theory in Theorems~\ref{thm:rho-null-fixedM} and \ref{thm:rho-null-fixedM-wass} into a practically usable inference procedure, avoiding the need to estimate the weighted chi-squared spectrum ${\gamma_\ell}$ in each application. This is especially valuable in the Wasserstein case, where the limit law is driven by an operator on $H^M$ rather than by a finite-dimensional matrix. 
    Second, the same resampling principle applies to both finite-dimensional Riemannian manifolds and one-dimensional Wasserstein responses, while placing no restriction on the predictor space beyond the partition geometry. In this way, the bootstrap extends FCC from an asymptotic effect-size characterization to an operational testing framework for directional Fr\'echet-mean dependence across distinct metric spaces.
\end{remark}

\section{Detailed simulation settings}\label{sec:sim-details}
This section records the exact numerical settings used for the figures in Section~\ref{sec:simulation} of the main text, which studies the empirical power of FCC in one Euclidean benchmark and four non-Euclidean settings. 
Power is estimated from $200$ Monte Carlo replications with $2000$ resampling replicates per replication; FCC uses fixed-$M$ wild-bootstrap replicates, while Ball and energy use permutations.
Note that for FCC computation, we use a $H$-packing partitioning strategy. Specifically, for each predictor sample we construct a nearest-prototype partition of the predictor space by first selecting $H$ prototype observations through a farthest-point rule and then assigning each observation to its nearest prototype under the predictor metric. 
This approach  yields a deterministic metric-based partition with controlled complexity. 
However, other partition constructions could also work in our framework.

In each model, we generate i.i.d. pairs $\{(X_i,Y_i)\}_{i=1}^n$ in the relevant metric space and control the dependence through a scalar parameter $\delta\in[0,1]$. We use $\delta$ here to denote the data-generating signal strength. When $\delta=0$, the construction reduces to independence; as $\delta$ increases, the dependence becomes stronger through a structured and generally nonlinear mechanism. Distances are computed using the natural metric for each object type: Euclidean distance on $\mathbb{R}^p$, chordal distance on the sphere $\mathbb{S}^{p-1}$ induced by the ambient Euclidean norm, the Log-Cholesky metric on $\mathcal{S}_p^+$, and the 2-Wasserstein metric on $\mathcal{W}_2$ implemented through quantile functions on a fixed grid.

\textbf{Setting I: $X,\ Y\in \mathbb{R}^p$}. Consider a sparse Euclidean vector model, set $X_i=(X_{i1},\dots,X_{ip})^\top\sim N(0,I_p)$ and generate $Y_i=(Y_{i1},\dots,Y_{ip})^\top$ by
\[
Y_{i1}=\delta\,\log\!\big(4X_{i1}^2\big)+0.8\,\varepsilon_i,\qquad \varepsilon_i\sim N(0,1),
\]
and for $j=2,\dots,p$, set $Y_{ij}\stackrel{iid}{\sim}N(0,1)$ independently of $(X_i,\varepsilon_i)$. This yields a sparse nonlinear dependence in which only the first coordinate pair carries signal, while the remaining coordinates are Gaussian noise. For Figure~\ref{fig:sim1_scalar}, we take $p=3$, $n\in\{20,40,60,80,100,150,200\}$, and $\delta=0.5$. 
Power is estimated from $100$ Monte Carlo replications with $1000$ resampling replicates per replication; FCC uses fixed-$M$ wild-bootstrap replicates, while the benchmark tests use permutations.
For FCC, we use $H=30$ and a minimum cell size $4$.

\textbf{Setting II:  $X,\ Y\in \mathbb{S}^2$}.
Generate $(X_i,Y_i)$ as unit vectors on the sphere $\mathbb{S}^2\subset\mathbb{R}^3$.
Let $\theta_i\stackrel{iid}{\sim}\mathrm{Unif}(-\pi,\pi)$ and set the mean direction of $X_i$ as
\[
\mu_X(\theta_i)=\big(\cos\theta_i,\ \sin\theta_i,\ 0\big).
\]
Define $\phi_i=\pi|\sin(k\theta_i)|$ for some $k\ge 2$ and specify the mean direction of $Y_i$ by
\[
\mu_Y(\theta_i)=\big(\cos(\delta\phi_i),\ 0,\ \sin(\delta\phi_i)\big).
\]
Add isotropic Gaussian noise in $\mathbb{R}^3$ and renormalize:
\[
X_i=\frac{\mu_X(\theta_i)+\sigma_X\xi_i}{\|\mu_X(\theta_i)+\sigma_X\xi_i\|},\qquad
Y_i=\frac{\mu_Y(\theta_i)+\sigma_Y\zeta_i}{\|\mu_Y(\theta_i)+\sigma_Y\zeta_i\|},
\]
where $\xi_i,\zeta_i\stackrel{iid}{\sim}N(0,I_3)$ and are independent of $\theta_i$.
Note that $\phi$ is many-to-one, so distant phases can yield similar responses. This produces a folded and non-injective dependence on $\mathbb{S}^{2}$ that is challenging for purely global distance-based summaries.
In Figure~\ref{fig:sim1}(a), we set $\delta=0.5$ and the sample size $n\in\{50,80,100,150\}$. 
For FCC, we use $H=15$ and minimum cell size $5$.

\textbf{Setting III:  $X,\ Y\in \mathcal{W}_2$}. 
We represent each distribution by its quantile function evaluated on a grid $q_\ell\in[0.01,0.99]$ for $\ell\in [m]$, and use the 2-Wasserstein distance
\[
	d_{\mathcal W}(x,y)^2=\int_0^1\{Q_x(u)-Q_y(u)\}^2\,du,
\]
approximated numerically on the grid. Let $U_i\sim\mathrm{Unif}(0,1)$ and let $Q_0(q)=\Phi^{-1}(q)$.
Define
\[
	Q_{X_i}(q)=U_i+\sigma_X Q_0(q),\quad  Q_{Y_i}(q)=\delta\sin(2\pi kU_i)+\sigma_Y \exp(\eta Z_i)\,Q_0(q),
\]
where $Z_i\sim N(0,1)$ is independent of $U_i$. 
Here $Y$ combines a periodic location signal in $U_i$ with sample-specific scale heterogeneity driven by $Z_i$, producing a nonlinear dependence structure in an infinite-dimensional metric space. 
In Figure~\ref{fig:sim1}(b), we use $\delta_X=1$ and $\delta_Y=0.4$. FCC uses $H=15$ and minimum cell size $5$.

\textbf{Setting IV:  $X,\ Y\in \mathcal{S}_p^+$}.
We generate $(X_i,Y_i)$ as SPD-valued pairs using Log-Cholesky coordinates.
Let $d=p+p(p-1)/2$ be the dimension of this parameterization.
For $p\ge 3$, choose two distinct off-diagonal coordinates corresponding to $L_{21}$ and $L_{31}$, denoted by $\mathrm{off1}$ and $\mathrm{off2}$.
Draw $U_i \stackrel{iid}{\sim} N(0,1)$ and define the signal $h(U_i)=|U_i|-\sqrt{2/\pi}$.
Construct coordinate vectors $V_{X,i},V_{Y,i}\in\mathbb R^d$ by
\[
(V_{X,i})_1 = U_i,\quad (V_{X,i})_{\mathrm{off1}}=0.6 U_i,\quad (V_{Y,i})_2=\delta h(U_i),\quad (V_{Y,i})_{\mathrm{off2}}=0.6\delta h(U_i),
\]
then add Gaussian perturbations 
$$V_{X,i}\leftarrow V_{X,i}+\sigma_X\epsilon^X_i,\quad V_{Y,i}\leftarrow V_{Y,i}+\sigma_Y\epsilon^Y_i.$$
To allow additional variation unrelated to the signal, we optionally add nuisance noise $\tau_{\text{nuis}}$ to the remaining coordinates of $V_{Y,i}$.
Finally, map each $V$ to an SPD matrix by forming a Cholesky factor $L$ with $\mathrm{diag}(L)=\exp(\text{log-diagonal entries})$ and setting $X_i=L_{X,i}L_{X,i}^\top, Y_i=L_{Y,i}L_{Y,i}^\top$. Distances are computed using the Log-Cholesky metric.
In this model, $X$ varies monotonically with $U$, while $Y$ depends on the transform $h(U)$, producing a non-injective dependence in $\mathcal{S}_p^{+}$ controlled by $\delta$.
Figure~\ref{fig:sim1}(c) choses $p=4$, $n\in\{50,80,100,150\}$, and $\delta=0.5$. For FCC, we use $H=15$ and a minimum cell size of $5$.

\textbf{Setting V:  $X,\ Y\in \mathcal{S}_p^+$}. This setting is designed to illustrate a dependent but FCC-null Wishart benchmark.
Let $\Sigma_0\in\mathcal{S}_p^+$ be a fixed baseline scale matrix, for example, the Toeplitz matrix with entries $(\Sigma_0)_{jk}=0.3^{|j-k|}$. Draw $U_i\stackrel{iid}{\sim}N(0,1)$ and define
\[
\Sigma_{X,i}=\exp(0.5U_i)\Sigma_0,\qquad
\Sigma_{Y,i}=\exp(0.5\delta U_i)\Sigma_0.
\]
Conditional on $U_i$, generate $X_i$ and $Y_i$ independently as Wishart random matrices with a common degree of freedom parameter $\nu>p-1$,
\[
X_i\mid U_i \sim W_p(\nu,\Sigma_{X,i}),\qquad
Y_i\mid U_i \sim W_p(\nu,\Sigma_{Y,i}).
\]
Distances are computed using the Log-Cholesky metric. When $\delta=0$, the distribution of $Y_i$ no longer depends on $U_i$, so $Y$ is independent of $X$. As $\delta$ increases, both $X$ and $Y$ respond to the same latent scale factor, producing a simple SPD dependence structure through paired Wishart scale matrices.
Figure~\ref{fig:sim1}(d) sets $p=4$, $\nu=16$, and $\delta=0.5$.

\bibliographystyle{plainnat}
\bibliography{ref}

\end{document}